\documentclass[twocolumn,showpacs,preprintnumbers,amsmath,amssymb]{revtex4}

\usepackage{graphicx}
\usepackage{dcolumn}
\usepackage{bm}

\newcommand{\be}{\begin{eqnarray}}
\newcommand{\ee}{\end{eqnarray}}
\newcommand{\ben}{\begin{eqnarray*}}
\newcommand{\een}{\end{eqnarray*}}
\newcommand{\la}{\left\langle}
\newcommand{\ra}{\right\rangle}

\newcommand{\half}{\frac{1}{2}}

\def\mb#1       {\mbox{\boldmath $#1$}}
\def\lab#1      {\hbox{\small #1} }

\begin{document}

\preprint{LSU-HETh-103B-2006}

\title{Model independent approach to studies of the confining dual Abrikosov vortex in $SU(2)$ lattice gauge theory}

\author{Richard W. Haymaker}
 \email{haymaker@rouge.phys.lsu.edu}
 \affiliation{Department of Physics and Astronomy, 
 Louisiana State University, Baton Rouge, Louisiana 70803-4001, USA}

\author{Takayuki Matsuki}
 \email{matsuki@tokyo-kasei.ac.jp}
\affiliation{Tokyo Kasei University, 1-18-1 Kaga Itabashi,  Tokyo 173-8602, Japan}

\date{July 20, 2006}

\begin{abstract}

We address the problem of determining the type I, type II or borderline dual 
superconductor behavior in maximal Abelian gauge  $SU(2)$  through the study of 
the dual Abrikosov vortex.  We find that significant electric currents in the simulation data call into question
the use of the dual Ginzburg Landau Higgs model in interpreting the data.   Further, two definitions of the penetration depth parameter take two different values.  The splitting of this parameter into two is intricately connected to the existence of electric currents.   It is important in our approach that we employ definitions of flux and electric and magnetic currents that respect Maxwell equations exactly for lattice averages independent of lattice spacings.  
Applied to specific Wilson loop sizes, our conclusions differ from those that use the dual GLH model.

\end{abstract}

\pacs{11.15.Ha, 12.38.Aw, 12.38.Gc}%

\maketitle

\section{Introduction}

In the search for simplicity in the physics of color confinement,  lattice
gauge theory models have been
studied extensively for a clue to a mechanism or an underlying principle
governing the phenomenon.   Spontaneous gauge symmetry breaking (SGSB) of the dual
$U(1)$, and the resulting  condensation of $U(1)$ monopole currents, defined after appropriate gauge fixing, remains a candidate.  The persistent monopole
currents of dual superconductivity in  pure $U(1)$ lattice models leads to
confinement of charge.   There has been some success in the postulate of
Abelian dominance in correlating monopoles and confinement physics but no
breakthrough in uncovering a definitive mechanism.
There are a number of reviews of the subject\cite{suzuki,cp,haymaker1,
polikarpov,digiacomo,cgpv,haymaker2,cgpz,ripka}.
Some of the principal directions include accounting for string tension in Abelian Wilson loops\cite{klsw,sy}, similarly for monopole dominance of Wilson loops\cite{snw,ss},  correlating percolating monopole clusters and 
confinement\cite{ht,bcgp,cz,ciks}, spontaneous  gauge symmetry breaking (SGSB) of dual $U(1)$ symmetry of the vacuum\cite{pisa3,pisa2,pisa1},  the dual Ginzburg Landau Higgs (GLH) 
model\cite{suganuma1,suganuma4} and studies of the structure of the dual Abrikosov vortex in the confining 
string\cite{sbh,mes,ph,cc,bss,bali,gips,kkis,kkisp} modeled by the dual GLH model.

Truncation to relevant variables invariably leads to systematic errors.  Ambiguities due the requirement of fixing the gauge \cite{bcp,mioys,ekmosy,sijs,fkst}  and Gribov 
copies\cite{bbms} contribute to these perennial problems.   Further it is difficult to see how this mechanism alone can explain all aspects of color confinement in arbitrary systems.  Nevertheless one can argue that the well established phenomena can be part of a larger picture.    

One possibility is that we are seeing SGSB as a general principle rather than just a confining mechanism.  The Pisa group\cite{pisa3,pisa2,pisa1} has given extensive evidence for this through QCD vacuum studies of the deconfining transition.  SGSB also manifests itself  in the existence of a
dual Abrikosov vortex in the confining string\cite{sbh,mes,ph,cc,bss,bali,gips,kkis,kkisp}; a complementary approach giving rather direct evidence for the same principle which is the subject of this paper.  

The dual Ginzburg Landau Higgs model has provided much insight into the physics 
of dual superconductivity\cite{sbh,mes,ph,cc,bss,bali,suganuma1,suganuma4,gips,kkis,kkisp}.   Bali et al.\cite{bss,bali} presented the first large lattice simulation of the dual Abrikosov vortex adding credibility to this approach to dual superconductivity. They employed techniques to select the best Gribov copy\cite{bbms}.   In addition to improving the gauge fixing, this technique reduces noise coming from the randomness of false maxima.  Gubarev et al.\cite{gips} found improvement over the fit of Bali et al. by using the numerical solution of a lattice version of the dual GLH model.  They concluded that dual superconductivity type lay on the borderline of type I and type II.  Koma et al.\cite{kkis,kkisp} 
went further and checked that the results obey scaling.  They also argue for a decomposition of the vortex into a photon part and monopole part.  Whereas previous work used a two dimensional model of the vortex, Koma et al. fit to a numerical solution of a three dimensional lattice model.  Both Bali et al. and Koma et al. concluded the theory is described by weakly type I dual superconductivity.

We present results showing that the strengths of the dual GLH 
model are accompanied by weaknesses when applied to the interpretation of the simulation of this theory in the standard simulation window.  Namely, there are identifiable parameters in the simulation taking different values that are equal by default in the dual GLH model.   This will inevitably increase systematic errors.
Also, in addition to the well-known area effect of confining magnetic currents, we show there is a significant perimeter effect of circulating electric currents which is absent in the model.  Although these electric current effects are expected to vanish for Wilson loop sources of very large spacial and temporal sizes, we show they can have an impact on the determination of the type I vs. type II dual superconductivity for fixed loop sizes.  The added parameter and the electric currents are intricately related.

In order to explore these issues we present an analysis that does not depend on the details of the dual GLH model.   

We regard the question of type I, type II or borderline to be the most important issue in these studies and we believe it can not be answered using analysis based on the dual GLH model.   Perhaps the best outcome would be if Gubarev et al.\cite{gips} are correct that the system is on the borderline between type I and type II.  We find it likely that type II behavior  goes away as one goes to larger quark separations and better suppression of excited states. But this is by no means definitive.   Type I brings with it special problems which are difficult to address as we will explain. 

Since the primary goal is to describe the total electric flux and the profile of the vortex as well as possible we consider it a top priority that our analysis satisfies Maxwell equations exactly for lattice averages.  This removes the arbitrariness in the choice of lattice  operators.

With this background, there are in our opinion  a number of reasons  
to revisit the dual vortex problem.  Let us sketch here the main issues presented in this paper:

{\bf (i) London relation in superconductivity} 

It is the lure of a general principle that draws us to further studies of the dual Abrikosov vortex in this paper.  
Let us first recall how this works in superconductivity\cite{tinkham} since this is the guide to the analysis of the simulation data.
Consider an infinitely long static Abrikosov vortex. Recall
\ben
\mb{J} ^{(e)} =  e \Im  \phi^* (\nabla - i e \mb{A} ) \phi ,
\een
where $\phi$ is the complex superconductivity order parameter.  Further consider SGSB  of $U(1)$ via
\be
\phi = v e^{i e \omega (x)}; \quad v \lab{ constant} .
\label{sgsb}
\ee
Then 
\ben
\mb{J} ^{(e)} =  - e^2 v^2 ( \mb{A} - \nabla \omega ) 
\een
and the curl of this gives the London  relation 
\be
\mb{B} = - \lambda^2  \lab{curl} \mb{J} ^{(e)},  \quad \lambda \equiv 1/ev .
\label{tate}
\ee
Equation (\ref{sgsb})  is the case for an extreme type II superconductor but it is also valid deep inside a type II 
superconductor far from a boundary where the complex phase degree of freedom of the order parameter dominates over the modulus.  (The order parameter is governed by a `Mexican hat' potential inside the superconductor and by a symmetric paraboloid potential in the normal material.  The spacial `kinetic energy' term excites the radial degree of freedom in the transition region.)

To construct a vortex with one unit of quantized magnetic flux assume that the phase of the order parameter increases by $2 \pi$ as the azimuthal angle $\varphi$ makes one complete path around the vortex at large  transverse distances where the current is  exponentially small.  Then
\be
\oint \mb{A} \cdot d \ell  &=&  \oint \nabla \omega \cdot d \ell, \\
\Phi_m =  \int \mb{B} \cdot d\mb{S} &=& \frac{2 \pi}{e}.
\label{fq}
\ee
A source consisting of a hypothetical Dirac monopole-antimonopole pair would generate an Abrikosov vortex between them containing  one unit of quantized magnetic flux, $\Phi_m$, which is equal to the monopole magnetic charge.  All the magnetic flux from the magnetic monopoles is accounted for in the vortex.  

To complete the picture using Ampere's law, Eq.(\ref{tate}) becomes
\be
\mb{B}  = \lambda^2 
\left\{ 
( \partial_x^2 + \partial_y^2 ) \mb{B} -\nabla( \nabla  \cdot \mb{B}) 
\right\} .
\label{diffusion}
\ee

There are two important aspects of this that we wish to note.  First, since  $\nabla \cdot \mb{B} = \rho_m = 0 $,  the square root of the proportionality factor in the London relation  is clearly the penetration depth in the superconducting material.   The Ginzburg-Landau model describes real materials hence has  vanishing magnetic charge.  In the four dimensional version, the GLH model, this is also true.  

Second is that the exterior solution to Eq. (\ref{diffusion}) is the Bessel function $K_0(r/\lambda)$.  This solution is applicable as long as the superconducting order parameter ramps up from the normal to the superconducting asymptotic value on a shorter length scale than the penetration depth, i.e. type II.  Then sufficiently deep in the superconducting medium the tail of the penetrating flux is described by Eq.(\ref{diffusion}). 

However if the penetration depth is shorter than the characteristic ramp-up length, i.e. type I, then there is no regime for which the $K_0(r/\lambda)$ solution is applicable.  There is not an easy fix.  The complication here is that the surface energy between the superconducting and normal material flips sign\cite{footnote1}  and an instability develops.  It is energetically favorable to make surfaces and the behavior of the penetrating flux changes dramatically\cite{huebener}
  See Tinkham\cite{tinkham} for a thorough treatment of the very different physics of the type I `intermediate state' and the type II `mixed state.' The  solutions of the GL model for  type I  superconductivity  in which magnetic flux penetrates  material involves fractals on many length scales\cite{callaway}.

{\bf (ii) Exact electric Maxwell equations} 

The above picture  translates nicely to a $U(1)$ lattice gauge theory in which a charge-anticharge pair arising from an Abelian Wilson loop produces a dual Abrikosov vortex.  More specifically, Zach, Faber, Kainz and Skala (ZFKS)\cite{zfks}, using a lattice Ward-Takahashi identity,  found a specific lattice operator for electric flux giving the exact {\em electric} Maxwell equation for lattice averages independent of lattice spacing $a$. For example in this formulation the divergence of the electric field is zero everywhere except on the Wilson line sources where it is equal to  the charge $e$ (see Sec. II B ).  Therefore using the ZFKS definition the total flux in the vortex, $\Phi_e$, is determined exactly by the charge on the Wilson loop in the lattice system which is of course the  gauge coupling $e$.  

In 1998, DiCecio, Hart and Haymaker (DHH)\cite{dhh}, generalized the ZFKS result to $SU(2)$ in the maximal Abelian gauge.   
This case is complicated by the existence of dynamical electric currents in addition to the static source. The extra
contributions come from the doubly charged vector matter fields, from the effects of gauge fixing and from ghosts.   Nevertheless by using the DHH definition, the total flux, $\Phi_{SU(2)}$ in the vortex is again an exact reflection of all  contributions to the charge distribution.  

We consider three definitions of flux in this paper, ZFKS, DHH and DeGrand-Touissant\cite{dt} (DT) and find that to a good approximation they give the same vortex profiles, but differ by a constant scale factor.  Bali et al.\cite{bss} 
did not state their definition but since they used the DT construction for the magnetic monopoles, it is reasonable to assume they used the same definition for the electric flux profile.  Koma et al.\cite{kkis} used the DT definition throughout.  For our DHH definition, and for $\beta = 2.5115$ we find the scale factor differs by $40\%$ from the DT definition.

{\bf (iii) Consistent magnetic Maxwell equations} 

One can then also get a consistent  exact {\em magnetic} Maxwell equation by 
adopting the same definition of flux in defining
the magnetic current\cite{mh}.   This is not the conventional procedure.   
The conventional one is the DT construction which identifies cubes with quantized
monopoles.  By using the DHH flux instead, the current is conserved but is not quantized in cubes.  Rather it is
smeared out among neighboring cubes.   

The magnetic Maxwell equations determine the solenoidal persistent magnetic currents that determine the shape of the vortex. 

These deviations from the more standard definitions are clearly tailored for this vortex problem.  The lattice
approach allows an infinite variety of definitions as long as they approach the same continuum limit. We argue in Section II how the DHH formulation might approach the standard  discrete monopole picture in the continuum limit.  However the DT 
definition is perhaps  the only one that can describe  percolating clusters\cite{ht,cz,ciks,bcgp} since their definition depends on the discretization of magnetic charge.  We use the DT definition here for the dual vortex problem as a digression from the main body of this work.  We see that it gives support to the truncation that retains only the percolating cluster.

The three definitions of flux considered in this paper are applied to the magnetic current and again to a good approximation give the same vortex profiles  but  differ from each other by a constant scale factor.

{\bf (iv) Model independent analysis}  

We chose to analyze the data in such a way as to test the  SGSB of dual $U(1)$ independent of specifics of the dual GLH model.  The model commits to a quartic interaction.  And it has vanishing dynamical electric currents.  
For gauge fixed  $SU(2)$ we determine a value of $\Lambda_d$ which satisfies a London equation asymptotically in the tail of the profile giving direct evidence for SGSB of dual $U(1)$. That is, considering the vortex aligned along the $z$ axis
\be
E_z = - \Lambda_d^2 ( \lab{curl} \mb{J} ^{(m)} )_z.
\label{dualtate}
\ee
This argument does  not depend on specifics of the solutions of the non-linear equations of the GLH model.

Next consider the profile of the fluxoid\cite{footnote2} 
${\cal E}_z$
\be
{\cal E}_z = E_z  + \Lambda_d^2 (\lab{curl} \mb{J} ^{(m)} )_z . 
\label{dualfluxoid}
\ee
The leading decaying exponential behaviors $E_z, \lab{curl} \mb{J} _m$$ \sim e^{-r/\lambda_d}$, should cancel giving a sub-leading behavior  $E_z + \Lambda_d^2 \lab{curl} \mb{J} _m \sim e^{-r/\xi_d}$ .   In an extreme type II dual superconductor the fluxoid would vanish exactly except on the axis of the vortex where
the system is `normal', i.e the order parameter is zero.   In our case the order parameter ramps up over a finite distance and we 
see a second exponential behavior after cancelation which defines the length scale corresponding to the violation of the London relation, i.e. the dual coherence length   $\xi_d$\cite{footnote3}\cite{footnote4}.
In this way we can estimate the three parameters $\Lambda_d$,  $\lambda_d$ and 
$\xi_d$ as long as  $\kappa_d > 1$, i.e. type II, where
\cite{footnote4a}
\ben
\kappa_d \equiv \frac{\lambda_d}{\xi_d}.
\label{pogo}
\een

This analysis of estimating exponential behaviors is applicable only if $\lambda_d > \xi_d$ since only then can one associate the two exponential behaviors with the parameters $\lambda_d$  and $\xi_d $ properly.  In other words we must have type II dual superconductivity for these studies to be meaningful.   For type I, the identification of the penetration depth from the behavior $K_0(r/\lambda)$ breaks down because of the `intermediate state' which is  a  mixture of normal and superconducting material.  Further the concept of a fluxoid as a sub-leading behavior breaks down.  

It is our opinion that this limitation to type II is not unique to our approach for reasons mentioned in subsection (i) above.   If one models the system with the dual GLH model and concludes type I behavior,  one needs to show that the solutions used are capable of producing the serpentine surfaces\cite{huebener} of the complex `intermediate state'. Further one needs to address how the lattice quantum averages relate to chaotic classical solutions.

{\bf (v) Electric currents and  dual superconductivity}

In the dual GLH model electric currents vanish everywhere except at the location of the Wilson loop static source. (Recall that the GLH model has vanishing magnetic currents.)  The dual of Eq.(\ref{diffusion}) implies  then
$\lambda_d = \Lambda_d$.   In the simulation we find that there are non-vanishing electric currents and further
as one would then expect $\lambda_d \neq \Lambda_d$.  This argues that our analysis can not be based on the 
details of the dual GLH model if we wish to learn more about these effects.

We are reporting here  that the $(z,t)$ curl of the electric current is non-vanishing on the same plaquette where we measure the $z$ component of the electric field which is the same plaquette where we measure the $ (x,y) $ curl of the magnetic current, i.e. $(\lab{curl} \mb{J} ^{(m)})_z$, out to rather large transverse distances.
A very localized cloud around the sources was well-known 
e.g.\cite{bdh,dhh,bss}. But our results  mean that this perimeter effect  
is much more pervasive  than expected.  For sufficiently large $R$ and $T$ dimensions of the Wilson loop, the currents and fields approach constants in $z$ and $t$ in the transverse midplane, resulting in the vanishing of the  $(z,t)$ curl of the electric current.  But these currents are relevant in the simulation window and need to be understood.  Determination of dual superconductivity parameters can only be definitive if the data is interpreted with the correct model.

Even if the electric currents were to vanish, we point out that $\Lambda_d$ depends on the definition of flux relative to the definition of magnetic current through the London relation,  whereas $\lambda_d$ is obtained from a characteristic decay length independent of normalization.    In the dual GLH model these are of course equal and hence failure to be consistent in the two definitions will lead to a  compromised fit.   Hence we argue for particular definitions based on the physical principle of satisfying Maxwell equations.

In Section II we review  and compare three definitions of flux that have appeared in the literature.  We argue in favor of 
particular forms of the flux and current which will satisfy the Maxwell equations  for lattice averages.  We show how
the distinction of the two parameters  $\Lambda_d$ and $\lambda_d$ are related to the existence of electric currents in the vortex. The generalization of the dual of Eq.(\ref{diffusion}) to the four dimensional lattice problem is the key.  This relation is exact for lattice averages only with our choice of definitions of flux, i.e. ZFKS for $U(1)$ and DHH for 
$SU(2)$, respectively and corresponding electric and magnetic currents which respect Maxwell's equations.

The numerical results are given in Sec. III.  We compare the three definitions of flux. Then using our preferred definition of flux, we measure the total flux in the vortex.  We then implement the model independent analysis described in sub-section (iv). 
We find type II behavior for Wilson loop sizes where others\cite{kkis,bss} find type I behavior.  Although these differences may be the result of differences in our approaches, or in the details, we regard the dual GLH model as the central culprit in the case of Koma et al. which we discuss in Sec. III C.  

A comparison of our work with Koma et al.\cite{kkis} and Bali et al. \cite{bss} shows that our determination of $\lambda_d$ is essentially unaffected by the Gribov copy problem at $\beta= 2.5115$ which we discuss in Sec. III C. The other parameters could have a small dependence as reflected by the small Gribov copy effect reported in the tails of the profiles\cite{kkis}.  Bali et al. makes use of the London relation as a first estimate of the penetration depth as we do which allows an interesting comparison possible between the results of these two groups and our results.   

We give examples of profiles of the electric currents that are contrary to the dual GLH model.

Also as a digression from the main body of this paper, we show the effect of using DT definition of magnetic current compared to a truncated DT definition in which only the single dominant percolating cluster is included.   

Section IV gives our summary and conclusions.

\section{Three definitions of flux}
Let us consider three definitions of field strength or flux, all agreeing to lowest order 
in the lattice spacing $a$.   If we require
that both electric and magnetic Maxwell equations for lattice averages of flux and current be satisfied,
then the specific form of the action implies a unique definition of flux.  Using any of the alternative definitions
introduces contributions, non-leading in $a$,  that violate Maxwell's equations. Consistency would then be restored only by going to the continuum limit.  

As we will see in Sec. II D below it is trivial to get an exact magnetic Maxwell equation as an operator relation.  This is not possible for the electric Maxwell equations.  If one had an exact electric Maxwell equation on a particular configuration, updating one link can change a single plaquette in the electric divergence violating the equality. Whereas in the magnetic divergence, changing one link always changes two plaquettes in a compensating fashion.  Attaining exact electric Maxwell equations is not trivial: the flux definition depends on the form of the action and the equations hold only for lattice averages.

\subsection{DeGrand Toussaint\cite{dt}(DT): $\widehat{F}_{\mu \nu}^{(1)}$} 
\begin{figure}[h]
\begin{center}
\includegraphics[trim=0 50mm 0 40mm,scale=0.4]{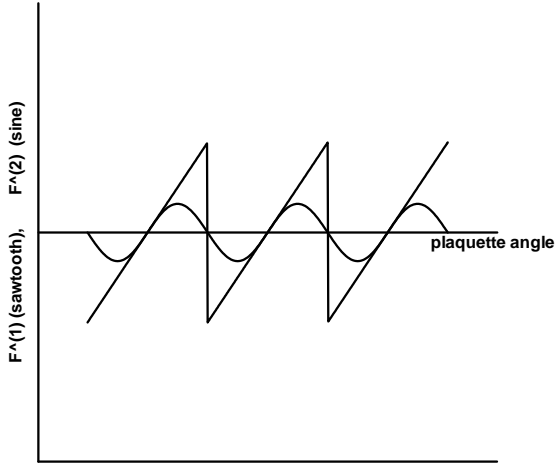}
\end{center}
\caption{$\widehat{F}_{\mu \nu}^{(1)}$ (sawtooth) and $\widehat{F}_{\mu \nu}^{(2)}$ (sine)
as a function of the plaquette angle $\theta_{\mu \nu}$}.
\label{venus}
\end{figure}

The first definition is that used by DeGrand and Tous-saint to define discrete
monopoles in the $U(1)$ theory:
\be
  \widehat{F}_{\mu \nu}^{(1)}(\mb{n} ) 
  &\equiv& \theta_{\mu \nu}(\mb{n} ) - 2 \pi n_{\mu \nu}(\mb{n} ), \label{maria}\\
 \theta_{\mu \nu}(\mb{n} ) &\equiv& \theta_\mu(\mb{n} )
-\theta_\mu(\mb{n} +\nu)  - \theta_\nu(\mb{n} ) +\theta_\nu(\mb{n} +\mu)\nonumber,
\ee
where $\theta_\mu$ refers to the $U(1)$ link angle in the domain
$-\pi < \theta_\mu < +\pi$.  The integers $n_{\mu \nu}$
are determined by requiring that  $-\pi < \widehat{F}_{\mu \nu}^{(1)} < +\pi$.
That is $\widehat{F}_{\mu \nu}^{(1)}$ is a periodic function of $\theta_{\mu \nu}$
with period $2 \pi$\cite{footnote5}.
We also refer to $\widehat{F}_{\mu \nu}^{(1)}$ as the `sawtooth' flux as shown in Fig.\ref{venus}.

\subsection{Zach-Faber-Kainz-Skala\cite{zfks}(ZFKS): $\widehat{F}_{\mu\nu}^{(2)}$ } 
A second definition  
\be
\widehat{F}_{\mu\nu}^{(2)}(\mb{n} ) &\equiv&\sin \theta_{\mu \nu}(\mb{n} ),
\label{betty}
\ee
has the property of giving the exact {\em electric} Maxwell equation for lattice averages for the case of the $U(1)$ gauge theory with Wilson action. 
Consider
\ben
Z_W(\epsilon_\mu(\mb{m} )) = \int [d \theta] \sin \theta_W
\exp\left( \beta S
\right),
\een
\be
S = \sum_{n,\;\mu > \nu} 
\left(
\cos \theta_{\mu \nu}(\mb{n} ) - 1\right)
,\;\;\;\;\;\;\;\;\;\;   \beta = \frac{1}{e^2}.
\label{capricorn}
\ee
The subscript of $Z_W(\epsilon_\mu(\mb{m} )) $ refers to the incorporation of the source into the partition function
and the argument is a variable defined as the shift of one particular link,
$\theta_\mu (\mb{m} ) \rightarrow \theta_\mu(\mb{m} ) +  \epsilon_\mu(\mb{m} ) $.
This translation
can be transformed away since the measure is invariant under such an operation.   
\be
\frac{\delta Z_W}{\delta \epsilon} = 0 \quad \Longleftrightarrow \quad
\beta \Delta_{\nu}^{-}
\la \widehat{F}^{(2)}_{\mu \nu} \ra_W
=
\la \widehat{J}^{(e)}_\mu \ra_W ,
\label{johnny}
\ee
where 
\ben
\la \cdots \ra_W = \frac{\la \cdots \sin\theta_W\ra}{\la \cos\theta_W\ra},
\een
and where difference operators are defined
\ben
\Delta_{\mu}^{+} \Psi(\mb{n} ) &\equiv&  \Psi(\mb{n}  +  \epsilon_\mu) - \Psi(\mb{n} ),\\
\Delta_{\mu}^{-} \Psi(\mb{n} ) &\equiv&   \Psi(\mb{n} ) - \Psi(\mb{n}  -  \epsilon_\mu),
\een
where  $\epsilon_\mu$ is a unit vector.

The  current $\la \widehat{J}^{(e)}_\mu \ra_W$ is that carried by the Wilson loop,
normalized to take values $-1,  0 , 1$, i.e. no factors of $e$.  (The expectation values are superfluous for this observable since the shifts produce these values directly from the Wilson loop factor in the partition function. The LHS of Eq. (\ref{johnny}) builds up these integer value from the lattice average.)

\subsection{DiCecio-Hart-Haymaker\cite{dhh}(DHH): $\widehat{F}^{(3)}_{\mu \nu}$ }

The third definition is applicable specifically to Abelian projected $SU(2)$. 
We restrict our attention to the maximal Abelian gauge defined as a  local maximum of
\ben
R &=&  \sum_{n,\mu}\lab{tr} \left\{\sigma_3 U_\mu({\bf n}) \sigma_3 U^{\dagger}_\mu({\bf n})  \right\},
\een
over the set of gauge transformations $\left\{g({\bf m}) = e^{i \alpha_i ({\bf m}) \sigma_i} \right\}$, 
$U \longrightarrow U^g$.  Taking $U$ to be the stationary value, the stationary condition is given by
\ben
F_{j {\bf n}}[U] &=& \left. \frac{\partial R[U^g]}{\partial \alpha_j({\bf n})}\right|_{\alpha = 0} = 0.
\een
The second derivatives entering in the Jacobian are given by
\ben
M_{j {\bf n}; i {\bf m}}(U) &=& \left. 
\frac{\partial^2 R[U^g]}{\partial \alpha_j({\bf n})\partial \alpha_i({\bf m})}\right|_{\alpha = 0} .
\een
The partition function is
\be
Z_W^{g.f.} (\epsilon_{\mu}^3(\mb{m} )) &=& \int [dU] \; 
\half Tr[ i\sigma_3 U_W (\mb{n} ) ] \times \nonumber \\&&
\exp\left( \beta S \right)
\prod_{j n}\delta(F_{j n}[U]) \; \Delta_{FP}, 
\label{su2-3}
\ee
where the Faddeev-Popov  Jacobian  is
\ben
\Delta_{FP} &=&  \lab{det} | M_{j {\bf n}; i {\bf m}} (U) |.
\een
An infinitesimal shift in this partition function has the added complication that it violates the
gauge condition.  This can be corrected by an infinitesimal accompanying gauge transformation.  Thus
the shift in one link affects all links.  However experience has shown that the effect drops off
rapidly with distance from the shifted link.  

Invariance under the shift leads to the definition of flux
\be
\widehat{F}^{(3)}_{\mu \nu}(\mb{n} ) &\equiv& 
C_\mu(\mb{n} ) 
C_\nu(\mb{n} + \mu)
C_\mu(\mb{n} + \nu) 
C_\nu(\mb{n} )\times \nonumber\\&&
  \sin \theta_{\mu \nu}(\mb{n} ),
\label{willie}
\ee
where the link variables are parameterized by 
$\theta_{\mu} (\mb{n} )$, $\phi_{\mu} (\mb{n} )$ and $\gamma_{\mu} (\mb{n} )$
\be
U_{\mu}(\mb{n} ) 
&=& 
\left( 
\begin{array}{cc}
C_{\mu}  e^{i \theta_{\mu} }  
& S_{\mu} e^{i (\gamma_{\mu}  - \theta_{\mu}  )}\\
- S_{\mu} e^{-i (\gamma_{\mu} - \theta_{\mu} )}
&
C_{\mu}  e^{-i \theta_{\mu} }
\end{array}
\right)
\label{bubba},
\ee
and where
\ben
C_{\mu} (\mb{n} ) &\equiv& \cos \phi_{\mu} (\mb{n} ),
\\
S_{\mu} (\mb{n} ) &\equiv& \sin \phi_{\mu} (\mb{n} ). 
\een

The decomposition of an $SU(2)$ link to $U(1)$ gives a $U(1)$ gauge link, $e^{i \theta_{\mu} }$, and a doubly charged matter field, $\sin \phi_{\mu} e^{i(\gamma_\mu - \theta_\mu)}$ transforming at the sites\cite{klsw,sy}.  
In the maximal Abelian gauge they also satisfy the vector field auxiliary condition making them bona fide charged vector matter fields as reviewed in Ref.\cite{bdh}.  

This form has the property of giving the exact {\em electric} Maxwell equation for 
lattice averages for this case of $SU(2)$ in the maximal Abelian gauge  with Wilson action
\be
\beta \Delta^-_\nu 
 \la    \widehat{F}^{(3)}_{\mu \nu}\ra
&=&
\la \widehat{J}_\mu^{(e) \lab{total} }\ra,
\label{jed}
\ee
where $\widehat{J}_\mu^{\lab{(e) total} }$ gets contributions from the
Abelian Wilson loop, the charged matter fields, gauge fixing and ghosts.
The normalization of the Wilson loop contribution to the current in this expression is analogous to the ZFKS case i.e. no introduced factors of $e$ or $a$. See Ref.\cite{dhh} for details.

\subsection{Consistency with the magnetic Maxwell equation}
For the second and third cases we have a unique flux $\widehat{F}_{\mu \nu}^{(i)}$, for  $i = 2,3$,
by requiring an exact lattice {\em electric} Maxwell equation.  Given this definition of flux
the {\em magnetic} Maxwell equation is
\be
 -\frac{1}{2}\epsilon_{\mu\nu\rho\sigma}\Delta_\nu^{+}
  \widehat{F}_{\rho\sigma}^{(i)} &=&
  \widehat{J}_\mu^{(m)} \quad \quad i = 2,3.
\label{jenny}
\ee
which gives a unique definition of the magnetic current.  However 
the monopole current is usually taken from the DT definition
\ben
  \widehat{J}_\mu^{(m)} &=&
 -\frac{1}{2}\epsilon_{\mu\nu\rho\sigma}\Delta_\nu^{+}
  \widehat{F}_{\rho\sigma}^{(1)} .
\een
(This current is normalized to give monopoles with a flux of $ 2 \pi n $ where $n$ is integer.)
Hence if we use the conventional  $\widehat{F}^{(1)}$ to define the monopole current, and
$\widehat{F}^{(2)}$ or $\widehat{F}^{(3)}$ respectively for $U(1)$ and $SU(2)$ theories
to get an exact expression for flux in the confining string, then the magnetic Maxwell equation 
is violated.

The {\em electric} Maxwell equation determines the total electric flux in the confining string and the 
{\em magnetic} Maxwell equation determines the transverse profile through the solenoidal currents.   
The only way for the calculation to be consistent with both Maxwell equations is to 
relax the usual procedure using the DT monopole definition and instead use
$\widehat{F}^{(2)}$ or $\widehat{F}^{(3)}$ when defining magnetic currents for the $U(1)$ and $SU(2)$ cases,
respectively.

A simple configuration will help illustrate the difference between $\widehat{F}^{(1)}$ and $\widehat{F}^{(2)}$.
Consider a single 
DT monopole with equal flux out of the six faces of the cube (and a Dirac string extending out from any face).  
Then the ratio of the 
$\widehat{F}^{(2)}$ flux out of this
cube compared to the $\widehat{F}^{(1)}$ flux gives
\be
\frac{6 \sin (2 \pi/6)}{6 (2 \pi/6)} \approx 0.83.
\label{horace}
\ee

On a large surface the total flux is the same for the two definitions.  
Since charge is conserved, the balance is made up by magnetic charge in the neighboring cubes.
We interpret this to mean
that with  magnetic currents defined with $\widehat{F}^{(2)}$,  the discrete monopoles become 
smeared  but maintain the same total magnetic charge.

\subsection{Comparison}
Figure \ref{venus} shows a comparison between the first two  definitions. 
We plot $\widehat{F}_{\mu \nu}^{(1)}$
as a function of $\theta_{\mu \nu}$, giving a ``sawtooth" shape.
Monopoles occur as a consequence of $\theta_{\mu \nu}$  crossing the
sawtooth edge, giving a mismatch of $2 \pi$ in the flux out of a cube.  
The sine function, $\widehat{F}_{\mu \nu}^{(2)}$, has no such discreteness and so the 
notion of discrete Dirac strings and Dirac monopoles is absent.  

However
as one approaches the continuum limit the two definitions merge.  To see this we note that
with increasing $\beta$, the Boltzmann factor supresses the plaquette angle to be in a more restricted 
effective domain in the neighborhood of $0$ mod $2 \pi$.    But we see from Fig. \ref{venus}
that the sin function and sawtooth function become the same in this domain up to corrections of the order of $a^2$.
We expect both forms to give the standard Dirac picture in the continuum limit.

The $\widehat{F}_{\mu \nu}^{(3)}$ definition is a modification of $\widehat{F}_{\mu \nu}^{(2)}$
involving factors of $C_\mu$, the cosine of the 
matter fields in the Abelian projection of the $SU(2)$ variables.   
Poulis\cite{poulis}pointed out that  $C_\mu$ has small fluctuations and can be taken approximately as a constant, i.e.  
$\la C_\mu \times {\cal O}\ra \approx \la C_\mu \ra  \times \la {\cal O}\ra$, where ${\cal O}$ is a generic operator.    Hence to this approximation, we find 
\ben
\la \widehat{F}_{\mu \nu}^{(3)} W \ra \rightarrow   \la C_\mu \ra^4 \times \la \widehat{F}_{\mu \nu}^{(2)} W \ra
\een
(The first factor on the right cancels in the physical normalization of the flux as we show in the next subsection.)  We will show in Sec. III A that for $\beta = 2.5115$ this approximation has violations of the order of $10\%$.

Since $C_\mu \approx 1 + O(a^2)$ in the Maximal Abelian gauge the fluctuations of this factor are suppressed  by an extra power of $a$. 
As an indication of the relative fluctuations in $C_\mu$,  our measurements of field strength have errors at best of the order of $1.0\%$ whereas the error in $\la C_\mu\ra$ is $0.01\%$.   

\subsection{Physical normalization of the field strengths}
The physical dimension of the fields is obtained from the leading order of a small $a$ limit.  
The definition of electric charge in the $U(1)$ theory is straightforward and well-known.  
The Wilson loop carries the charge of the gauge coupling constant.

The Abelian projected $SU(2)$ case is not as clear cut.   
Measurement of electric flux at the site of the Wilson loop includes the bare charge. 
In addition there are contributions due to the dynamical charged matter fields,  the effects of gauge fixing and ghosts\cite{dhh}.  Further an electrically neutral combination of the matter fields can not be 
completely disentangled from the gauge fields.  We propose a normalization that involves the effects of 
these matter fields.

\subsubsection{$U(1)$ theory with Wilson Action}
Given the electric Maxwell equation, Eq.(\ref{johnny}) and the magnetic Maxwell equation, Eq.(\ref{jenny}) (for $i = 2$
and with the Wilson action, Eq.(\ref{capricorn}))
one arrives in the usual way that
\ben
\beta = \frac{1}{e^2} \quad \lab{and} \quad e_m = \frac{2 \pi}{e}
\een
As mentioned in Subsection B the current $\widehat{J}^{(e)}_\mu$ is normalized to unity on the Wilson loop.
Recall that $\widehat{J}_\mu^{(m)}$ is normalized to take the value $2 \pi$ for a DT monopole.
The argument in subsection  D above shows that the same normalization holds for a smeared monopole constructed 
from ZFKS flux.  The normalizations are
\be
  F^{(2)}_{\mu \nu}  &=&
\frac{1}{e a^2} \widehat{F}^{(2)}_{\mu \nu}, 
\label{welch1}
\\
J^{(e)}_\mu  &\equiv&  \frac{e}{a^3} \widehat{J}^{(e)}_\mu,
\label{welch2}
\\
J^{(m)}_\mu   &\equiv& \frac{2 \pi}{e}  \left( \frac{1}{2 \pi a^3}\widehat{J}^{(m)}_\mu \right).
\label{welch3}
\ee

\subsubsection{$SU(2)$ theory with Wilson Action in the maximal Abelian gauge}

Generalizing to this case we have
\be
\beta\Delta_{\nu}^{-}
\la \widehat{F}^{(3)}_{\mu \nu} \ra
&=&
 \la \widehat{J}^{(e) \lab{Abelian Wilson loop} }_\mu \ra\nonumber\\ &&
+ \la \widehat{K}^{(e) }_\mu \ra,
\label{barney}
\ee
where 
$\widehat{K}^{(e)}_\mu $ is the sum of the three dynamical terms in the current and
$\widehat{J}^{(e) \lab{Abelian Wilson loop} }_\mu $
is normalized to unity on the Wilson loop.  The expectation values are understood to be taken in the background of an Abelian Wilson loop source with gauge fixed configurations.

Using the notation of Sec. II C we write the Wilson action
\ben
\beta \sum_P \half \lab{Tr} (U_{\mu\nu}) 
&=& \beta  \sum_P 
C_\mu(\mb{m} ) 
C_\nu(\mb{m} + \mu)C_\mu(\mb{m} + \nu) \times \\ &&
 C_\nu(\mb{m} ) 
  \left\{\cos \theta_{\mu \nu}(\mb{m} ) - 1\right\}
+ \cdots
\een
For the maximal Abelian gauge we will make use of the fact that the fluctuations of $C_\mu$ are 
suppressed.  If we take $C_\mu$ to be constant and compare with the $U(1)$ action, Eq.(\ref{capricorn}),
we can introduce the $U(1)$ charge through the relation.
\be
\frac{1}{e^2} &=& \beta \la C_\mu \ra^4 .
\label{adam}
\ee
Then we can write
\ben
\frac{\Delta_{\nu}^{-}}{a}
\la \frac{\widehat{F}^{(3)}_{\mu \nu}}{e a^2 \la C_\mu\ra^4} \ra
&=&
e \frac{\la \widehat{J}^{(e) \lab{Abelian Wilson loop} }_\mu \ra}{a^3} \nonumber\\ &&
+ \la  e\frac{\widehat{K}^{(e) }_\mu}{a^3} \ra.
\een
In the simulation for $\beta = 2.5115$  we found $\la C_\mu \ra = 0.94784(4)$ which gives
\be
e = 0.7024(1).
\label{valueofe}
\ee
This is the charge carried by the bare Abelian Wilson loop.  Equation (\ref{valueofe}) is only
an estimate of the value of $e$ because $ C_\mu $ has small fluctuations and can not be considered strictly a constant.

Considering finally the magnetic equation, Eq.(\ref{jenny}) for $i = 3$ we arrive at a normalization 
analogous to that for the $U(1)$ case
\be
  F^{(3)}_{\mu \nu}  &=&
\frac{1}{e a^2 \la C_\mu \ra^4} \widehat{F}^{(3)}_{\mu \nu}, 
\label{grape1}
\\
J^{(e)}_\mu  &=&  \frac{e}{a^3} \widehat{J}^{(e)}_\mu,
\label{grape2}
\\
J^{(m)}_\mu   &=& \frac{1}{e a^3 \la C_\mu \ra^4}   \widehat{J}^{(m)}_\mu.
\label{grape3}
\ee
\subsection{Effects due to electric currents using  DHH definitions}

Consider the classical magnetic Maxwell equations
\be
J^{(m)}_\beta &=& \;
 -\half \epsilon_{\beta \gamma \mu \nu}
 \; \frac{1}{a}\Delta^+_\gamma F_{\mu \nu} .
\label{reggie}
\ee
Take the curl of this
\be
 -\epsilon_{\sigma \rho \alpha \beta} \Delta^-_\alpha J^{(m)}_\beta &=& \;
 \left\{ 
 \epsilon_{\sigma \rho \alpha \beta}
 \half \epsilon_{\beta \gamma \mu \nu}\right\} \frac{1}{a}
 \; \Delta^-_\alpha \Delta^+_\gamma F_{\mu \nu} \nonumber ,
\\
 &=& -\frac{1}{a}\left\{
\Delta^-_\alpha \Delta^+_\sigma F_{\alpha \rho} - \Delta^-_\alpha \Delta^+_\rho F_{\alpha \sigma}
  \right.\nonumber\\&& \left.
\quad + \Delta^-_\alpha \Delta^+_\alpha F_{\rho \sigma}
\right\}.
\label{marvin}
\ee
Using the classical electric Maxwell equation, 
\be
\frac{1}{a}\Delta^-_\alpha  F_{\rho \alpha} = J^{(e)}_\rho,
\label{bethe}
\ee
we get the following relation
\be
\epsilon_{\sigma \rho \alpha \beta}
\Delta^-_\alpha   J^{(m)}_\beta 
 = -
\left( \Delta^+_\sigma J^{(e)}_\rho
  -  \Delta^+_\rho  J^{(e)}_\sigma \right)
  +\Delta^-_\alpha \frac{1}{a}\Delta^+_\alpha  F_{\rho \sigma}.\;\;\;
\label{lawnandgarden}
\ee
Now using the DHH form in the two currents and the flux, i.e.  $F^{(3)}_{\mu \nu}$, this expression must hold 
for expectation values
\be
&&\epsilon_{\sigma \rho \alpha \beta}
\Delta^-_\alpha  \la J^{(m)}_\beta \ra
 = \nonumber\\
&&-\left( \Delta^+_\sigma \la J^{(e)}_\rho \ra
  -  \Delta^+_\rho \la J^{(e)}_\sigma  \ra \right)
  + \Delta^-_\alpha \frac{1}{a}\Delta^+_\alpha \la F^{(3)}_{\rho \sigma}\ra.\quad
\label{gardens}
\ee

Applying this to a vortex oriented along the $z$ axis and choosing $\sigma = 3$ and $\rho = 4$
\be
\frac{1}{a}
\left(
\Delta^-_1  \la J^{(m)}_2 \ra -  \Delta^-_2  \la J^{(m)}_1 \ra \right)  && \nonumber \\
-\frac{1}{a^2}\left(\Delta^-_1\Delta^+_1 + \Delta^-_2\Delta^+_2 \right) \la F^{(3)}_{4 3} \ra
 = &&  \nonumber \\
-\frac{1}{a}
\left( \Delta^+_3 \la J^{(e)}_4 \ra
  -  \Delta^+_4  \la J^{(e)}_3 \ra \right) && \nonumber\\
  + \frac{1}{a^2}\left(\Delta^-_3\Delta^+_3 + \Delta^-_4\Delta^+_4    \right) \la F^{(3)}_{4 3}\ra . &&
\label{steve}
\ee

The electric current can survive in the dual GLH model but only as a lattice 
artifact and it vanishes in the continuum limit.   Further the second derivative terms  on the RHS of 
Eq.(\ref{steve}) are designed to
be as small as possible by the choice of the source.  
Assume first that the RHS vanishes.  Then the London relation becomes
\ben
\la E_z \ra  + \Lambda^2_d \left(\la \lab{curl} \mb{J} ^{(m)} \ra\right)_z = \la E_z \ra - \lambda^2_d  \nabla^2_\perp  \la E_z \ra = 0.
\een
This clearly identifies $\Lambda^2_d = \lambda^2_d$ as the penetration depth in the dual superconductor. 

However the RHS does {\em not} vanish for lattice averages in our simulation as we show in Sec. III.  In the standard simulation window, the terms on the RHS can be comparable to the terms on the LHS.   
Hence the value $\Lambda_d $ as measured by the London relation in the tail of the profile does {\em not}
control the rate of transverse fall-off of the profile.    In a dual GLH model it does.  For this reason we
choose not to rely on a fit to the dual 
GLH model but concentrate instead on verifying the 
model-independent SGSB, and estimating the  three parameters $\Lambda_d $, $\lambda_d $, and $\xi_d$.

\section{Simulation}

Our measurements were on 208 gauge-fixed configurations on a $32^4$ lattice, with $\beta = 2.5115$.   Each update consisted of a $10$ hit metropolis sweep and an overrelaxation sweep.   We made $13$ runs on $16$ independent nodes.   D	ropping $2000$ thermalization updates (on each node),  we made measurements on every $100 th$ update. 

We gauge-fixed to the Maximal Abelian Gauge (MAG) using overrelaxation with the criterion of the average of the absolute value of the off-diagonal matrix element of the MAG adjoint operator $ < 10^{-6}$.

We measured Wilson loops in which spacial links were fattened  through 100 iterative steps by adding spacial staples of weight equal to the original link.  In a spot check of the data sets, increasing the number of iterations from 100 to 200 and/or  decreasing  the weighting factor of the staple contribution by one half had no appreciable effect.  Although the
fattening appears to have saturated, the data shows a residual $T/a$ dependence, where $T/a$ is the time extent of the
Wilson loop in lattice units.

All vortex profile graphs presented here  are transverse slices through the mid-plane, the $(x,y)$ plane, on the quark-antiquark axis, the $z$ axis.   
Fitted parameter values are calculated for $6$ quark-antiquark separations
 $R/a = 3,5,\ldots, 13$ and for $7$ time separation of $T/a = 3,4, \ldots, 9$.   We used Minuit for the fits and quote Minuit errors in the parameters determined by $\Delta \chi^2 = 1$.

We used $a^2 F^{(3)}_{\mu \nu}$ throughout for the definition of flux in order to give the correct electric Maxwell equations.   Similarly we used the same definition in constructing the magnetic current $ a^3 J^{(m)}_\mu$ in order to get the correct magnetic Maxwell equation.\  (Exceptions of course include graphs comparing different definitions and truncated vs. complete monopole loops.)

Noise in these data is a problem.   We reduced this by taking an azimuthal angular average over an annular region of width $= 1$ in lattice units in the transverse, $(x,y)$ plane, weighting the data by the fractional area overlap of the data point 
plaquette to the annulus of radius $r$
\be
\la \cdots \ra_\phi   &=& \frac{1}{2 \pi r}\int_{r < r' < (r+1)}(\cdots ) da' .
\label{annulus}
\ee
Using this together with Stokes theorem we have a convenient numerical evaluation of
\be
\oint_{r' = r} \mb{J} \cdot d \mb{r'}  &=& 2 \pi r \la J_\phi(r)  \ra_{\phi},
\nonumber \\  &=&  \int_{r' < r } (\mb{\lab{curl} J} ) \cdot d \mb{a'} , \nonumber\\
&=& -\int_{r < r'} (\mb{\lab{curl} J} ) \cdot d \mb{a'}  ,
\label{lineintegral}
\ee
where we used the identity $\int  (\mb{\lab{curl} J} ) \cdot d \mb{a'} = 0$.  

Since the current is a first derivative of the flux and the curl of the current is a second derivative, errors become
more difficult for the latter.  However by integrating an equation over a transverse area involving the curl we are
back to first derivatives which are more manageable.

Except where a fit to the data is noted, the lines in the graphs connect the data points.  All axes on all graphs are
dimensionless.  For physical normalizations, one can take the standard value $a = 0.086\lab{fm} = 0.44 GeV^{-1}$.
\begin{figure}[h]
\vskip 0.7in
\includegraphics[trim=20mm 0 0  0  ,scale=0.4]{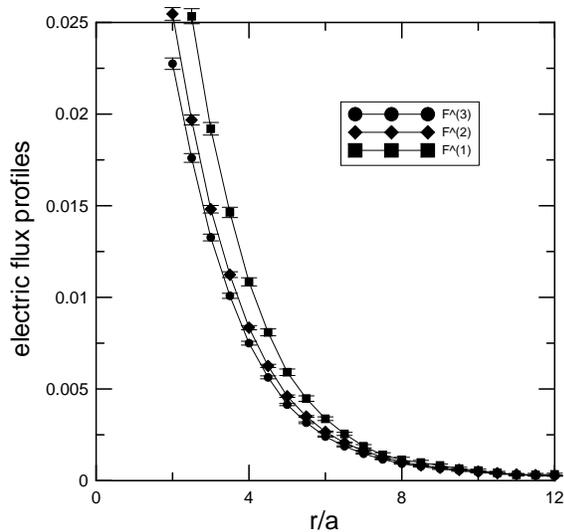}
\vskip -1.1in
\caption{Profiles for three definitions of the electric flux $a^2 F_{3 4}^{(1,2,3)}$ through the transverse
plane.  $R/a = 7$, $T/a = 3$, with azimuthal angular averaging.}
\label{pluto1}
\end{figure}
\begin{figure}[h]
\begin{center}
\vskip 0.7in
\includegraphics[trim=20mm 0 0 0,scale=0.4]{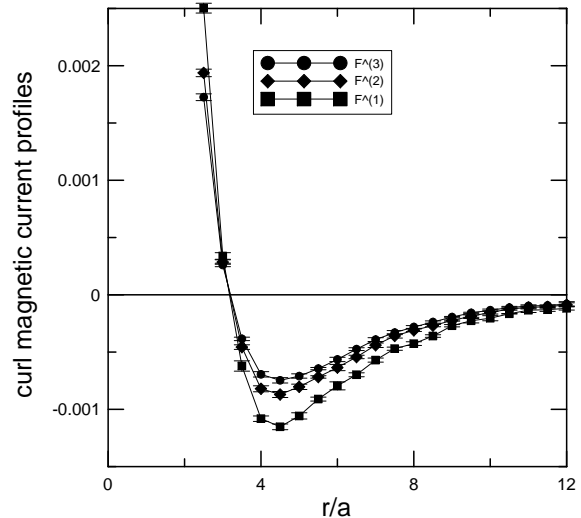}
\vskip -1.1in
\end{center}
\caption{
Same as Fig. \ref{pluto1} for the $z$ component of  $a^4\lab{curl} J^{(m)}$.} 
\label{pluto2}
\end{figure}
\begin{figure}[b]
\begin{center}
\vskip 0.7in
\includegraphics[trim=20mm 0 0 0,scale=0.4]{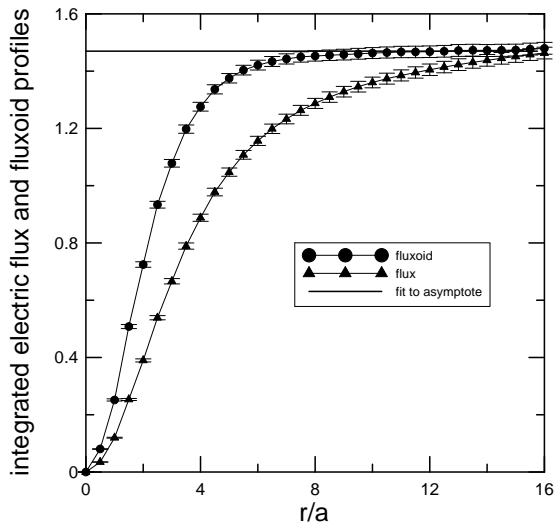}
\vskip -1.1in
\end{center}
\caption{Flux and fluxoid integrated over a disk of radius $r/a$.  $R/a = 7$, $T/a = 3$.}
\label{jupiter1}
\end{figure}
\begin{figure}[h]
\begin{center}
\vskip 0.7in
\includegraphics[trim=20mm 0 0 0,scale=0.4]{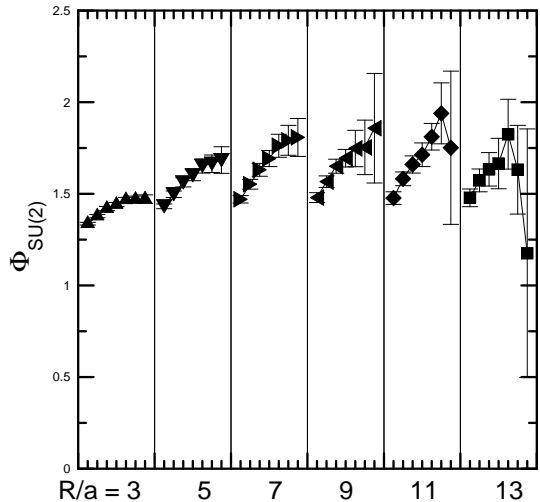}
\vskip -1.0in
\end{center}
\caption{Total electric flux in mid transverse plane. Within each group labeled by $R/a$, the $7$  ticks correspond to $T/a = 3,4, \ldots, 9$. }
\label{jupiter2}
\end{figure}

\begin{table*}
\begin{center}
\begin{tabular}{l l l l} \hline
& gauge action summand 
& gauge coupling 
& quantized vortex flux unit
\\ \hline
 GLH model
& $\beta \cos {\theta_{\mu \nu}}$ 
& $e = \beta^{-1/2}$ 
& $\Phi_m = e_m = 2 \pi/ e$
\\ \hline
dual GLH  model
& $\beta_{(d)} \cos {\theta^{(d)}_{\mu \nu}}$ 
& $e_m = \beta^{-1/2}_{(d)} $ 
& $\Phi_e = e = 2 \pi/ e_m$
\\ \hline
$U(1)$ theory
& $\beta_{U(1)}\cos {\theta_{\mu \nu}}$ 
& $e_{U(1)} = \beta_{U(1)}^{-1/2} $ 
& $\Phi_{U(1) } = e_{U(1)} $
\\ \hline
$SU(2)$ MAG theory \hspace{0.3cm}
& $\approx \beta_{SU(2)} \la C_\mu\ra^4 \cos {\theta_{\mu \nu}} + \cdots$ \hspace{0.3cm}
& $e_{SU(2)} = \beta_{SU(2)}^{-1/2} \la C_\mu\ra^{-2}$ \hspace{0.3cm}
& $\Phi_{SU(2)} = e_{SU(2)}  + \lab{(anti)screening} $
\\
& $\la C_\mu\ra = 0.94784(4)$
&
&
\\
& $\beta_{SU(2)}= 2.5115$
& $e_{SU(2)} = 0.7024(1)$
& $\Phi_{SU(2)} =  1.72(3)$
\\ \hline
\end{tabular}
\end{center}
\caption{Relationship between gauge coupling and quantized flux in the vortex. }
\end{table*}

From this point on, we suppress the expectation value bracket notation, leaving it understood that we are taking expectation values with a smeared Wilson loop.  All flux, and currents are constructed using 
 the DHH form $F^{(3)}_{\mu \nu}$ except in Sec. III A and III E.

\subsection{Comparing flux definitions} 
Recall the three definitions of flux: $F_{\mu \nu}^{(1,2,3)}$: Eq.(\ref{maria}, \ref{betty}, \ref{willie}).
Figure \ref{pluto1} shows the $E_z$ profile for the three definitions of flux.   
Surveying a large variety of quark separations $R = 3,5,\ldots, 13$ and time extents $T = 3,4, \ldots, 9$ the three 
definitions appear to differ only in the scale.  We did not notice any significant difference in shape.   The scale factors over this  range are approximately constant for $\beta = 2.5115$.
\be
 F_{\mu \nu}^{(1)} &\approx& 1.3  F_{\mu \nu}^{(2)},\nonumber  \\
 F_{\mu \nu}^{(2)} &\approx& 1.1  F_{\mu \nu}^{(3)}.
\label{comp}
\ee
We note that if the small fluctuations were in fact absent in $C_\mu$, Eq.(\ref{willie}), 
then $F_{\mu \nu}^{(3)} =  F_{\mu \nu}^{(2)}$ since  the $C_\mu$ factors cancel out in the normalization, 
Eq.(\ref{grape1}).

Figure \ref{pluto2} shows the $z$ component of curl $\mb{J} ^{(m)}$ constructed from the three definitions of flux. The same observations hold here as for the flux. Note that the connecting lines cross close to zero, further indicating just a scale factor.   They scale with the same factors as the flux to about $1\%$.   

Figures \ref{pluto1} and \ref{pluto2} show qualitatively the signal for a dual Abrikosov vortex no matter which 
definition is chosen.
For large transverse distance, e.g. in this case $r/a > 5$, the London relation holds, i.e.
the tails can be arranged to cancel, 
\be
\lab{fluxoid} : {\cal E}_z  \equiv E_z + \Lambda_d^2 \left(\lab{curl} \mb{J} ^{(m)}\right)_z  \approx 0.
\label{marslanding}
\ee
The integral of the 
$(\lab{curl} \mb{J} ^{(m)})_z $ over the whole plane vanishes hence the integrand must change sign.  For an extreme type II 
dual superconductor,  the profiles match and cancel everywhere except for a delta function contribution at $r=0$, giving the sign change needed for vanishing of the surface integral.  In our case the  scale defining the breakdown of the London relation is  the dual coherence length $\xi_d$. 

Figures \ref{pluto1} and \ref{pluto2} also show that by varying the relative definition of the two quantities one would change the value of $\Lambda_d$ without changing the  value of  $\lambda_d$  leading to systematic errors in the analysis in a model in which the two parameters are forced to take on the same value.

\subsection{Total electric flux}
We first consider the total flux integrated over the transverse mid plane since these data generally have the smallest errors over the range of the 42 data sets.

In Fig. \ref{jupiter1} we plot the integrated flux and fluxoid in the interior of the disk $r'\leq r$
\ben
\int_{r' \leq r} E_z \;\; da'; \quad \quad \int_{r' \leq r} {\cal E}_z \;\; da'.
\een
The integrated fluxoid reaches the asymptotic value at smaller $r/a$ than the integrated flux because it corresponds to the ramping up of the order parameter, not the penetration of the flux.  They must have the same asymptote because the surface integral of the fluxoid over the whole transverse plane vanishes.
\ben
\Phi_{SU(2)}  \equiv \int E_z \;\; da'  =  \int {\cal E}_z \;\; da'.
\een

Figure \ref{jupiter2} shows the total flux obtained from the radial asymptote of the fluxoid
as a function of $R/a$ and $T/a$.  This shows that for each $R/a$ the data goes to an asymptote for large $T/a$, at least for those values of $R/a$ where the errors do not mask the behavior.  
\begin{figure}[h]
\begin{center}
\vskip 1.0in
\includegraphics[trim=30mm 0 0 0,scale=0.4]{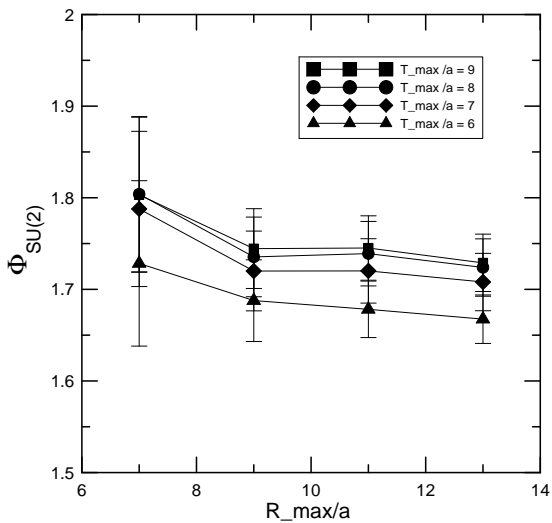}
\vskip -1.1in
\end{center}
\caption{Extrapolation to $T/a=\infty$ of  the integrated transverse flux based on data for spacial separations $ 3,5, \ldots R_{max} /a$ for 4 values of $T_{max}/a$.}  
\label{vulcan}
\end{figure}
With this as a guide we do an exponential fit to extract $\Phi_{SU(2)}$  for infinite $T/a$ values, based on one term in the transfer matrix.  
\ben
\Phi_{SU(2)}  &=& A + B e^{-C T/a}.
\een
Further we do a global fit, using all the data available to arrive at a value of total flux and we then test it for independence of $R$.  To do this we take the above form and allow an extra dependence linear in $R/a$ in the parameters $B$ and $C$ giving
\ben
\Phi_{SU(2)} &=& A + (B + B'R/a) e^{-(C + C'R/a)T/a}.
\een

The results are shown 
in Fig. \ref{vulcan}. 
We plot the value of $A$ as a function of the largest $R/a$ included in the fit, $R_{\lab{max} }/a$. 
Similarly, the family of curves shows the dependence on  $T_{\lab{max} } /a$.
The value of $\Phi_{SU(2)}$ determined this way using all 42 data sets is
\ben
\Phi_{SU(2)} = \int E_z \;\; da &=& 1.72(3), \quad \chi^2 /\lab{d.f.}  = 1.0  .
\een

As an alternative to this extrapolation, we can also just take a few data points from 
Fig. \ref{jupiter2} for values of $R/a$ and $T/a$ large where the value seems to be stable but small enough so that statistical errors are manageable: 
\ben
 T/a = 6
\left\{
\begin{array}{rr} 
R/a = 7 & \Phi_{SU(2)} = 1.76(6),\\
R/a = 9& \Phi_{SU(2)} = 1.75(9), \\
R/a = 11& \Phi_{SU(2)} = 1.81(9).
\end{array}
\right.
\een

Table I summarizes the relationship between the gauge coupling constant and the quantized vortex flux for the four cases
considered here.  For the $U(1)$  and $SU(2)$ cases, the Wilson loop carries the charge $e_{U(1)}$ and
$e_{SU(2)}$ respectively.  For the   $U(1)$ case, the quantized unit of flux is also $e_{U(1)}$ and so 
the elementary charge produces exactly one unit of quantized flux in the vortex.  In the $SU(2)$ case however there
is a dynamical charge distribution generated by the source  exhibited by the large value of 
$e_{\lab{dynamical} } \equiv   \Phi_{SU(2)} - e_{SU(2)} = 1.02(3)$.  (Since Gauss' law is satisfied
exactly in this formalism, $\Phi_{SU(2)}$ measures exactly the total charge on each side of the transverse
plane.) 

If we are correct that the proper interpretation of the simulation data implies that there is a dual gauge theory
operating, fixing the quantization of flux to one unit, then from lines $2$ and $4$ in Table I we conclude that the fundamental unit of flux in that dual theory is  $\Phi_{SU(2)} =  1.72(3)$, implying that the gauge coupling constant in that dual
theory is $e_m = 2 \pi/\Phi_{SU(2)}$.  

Is the assumption of one unit of flux correct?   
Could it be that the correct interpretation of our data  is a superposition of vortices with multiple units of flux?  

The $U(1)$ example suggests that this is indeed possible\cite{shb}.    Consider a lattice with a small extent in the  $z$ direction, i.e. along the alignment of the vortex.   Consider the case in which the distance between the sources is the same in both directions around the torus. And take a large enough time extent of the Wilson loop so that the spacial direction of the flux is not biased in one direction or the other. 
 We will obviously find half a unit of flux in each direction.    We are describing a system in which the vortex can occur in either of the two directions.   Our interpretation is that lattice average describes a vortex that can go in either direction and that the measurement on one side is a superposition of $0$ and $1$ units of flux. 

Therefore since a superposition is obviously possible, we think it is best to regard  $e_m = 2 \pi/\Phi_{SU(2)}$ as a tentative result subject to a better understanding of the dual GLH model or a generalization of it to include the electric currents.

We wish to note some interesting aspects of Fig. \ref{jupiter2}.  For each value of $R/a$, $\Phi_{SU(2)}$ increases with $T/a$.  The Wilson loop carries an electric current and the dynamical currents circulate in the same sense. This is evident when looking at a fixed time slice since there is antiscreening in the divergence of the current. Since increasing $T/a$ suppresses excited states, we must conclude that the electric current associated with them circulates in the opposite direction than in the ground state.   Further we note that the saturation of the anti-screening cloud as a function of $R/a$ is also evident at fixed $T/a$.  We take the case $T/a = 7$ which has manageable errors for increasing $R/a$.  In that case the anti-screening  already appears to be saturated for $R/a = 5$.

\subsection{Three parameters of the vortex} 
\begin{figure}[h]
\begin{center}
\vskip 0.7in
\includegraphics[trim=20mm 0 0 0,scale=0.4]{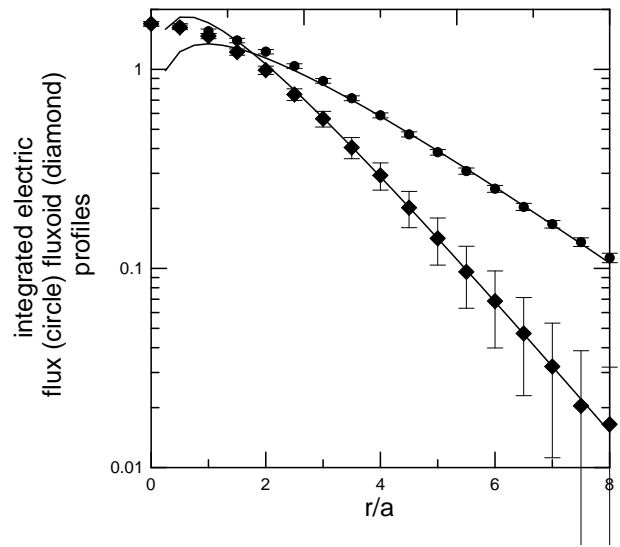}
\vskip -1.1in
\end{center}
\caption{Profiles of electric flux and fluxoid integrated outside a disk of radius $r/a$. Fits 
corresponding to  Eqs. (\ref{katrina}) and (\ref{rita}) also shown. $R/a = 7$, $T/a = 6$.}
\label{mars1}
\end{figure}
\begin{figure}[h]
\begin{center}
\vskip 1.0in
\includegraphics[trim=20mm 0 0 0,scale=0.4]{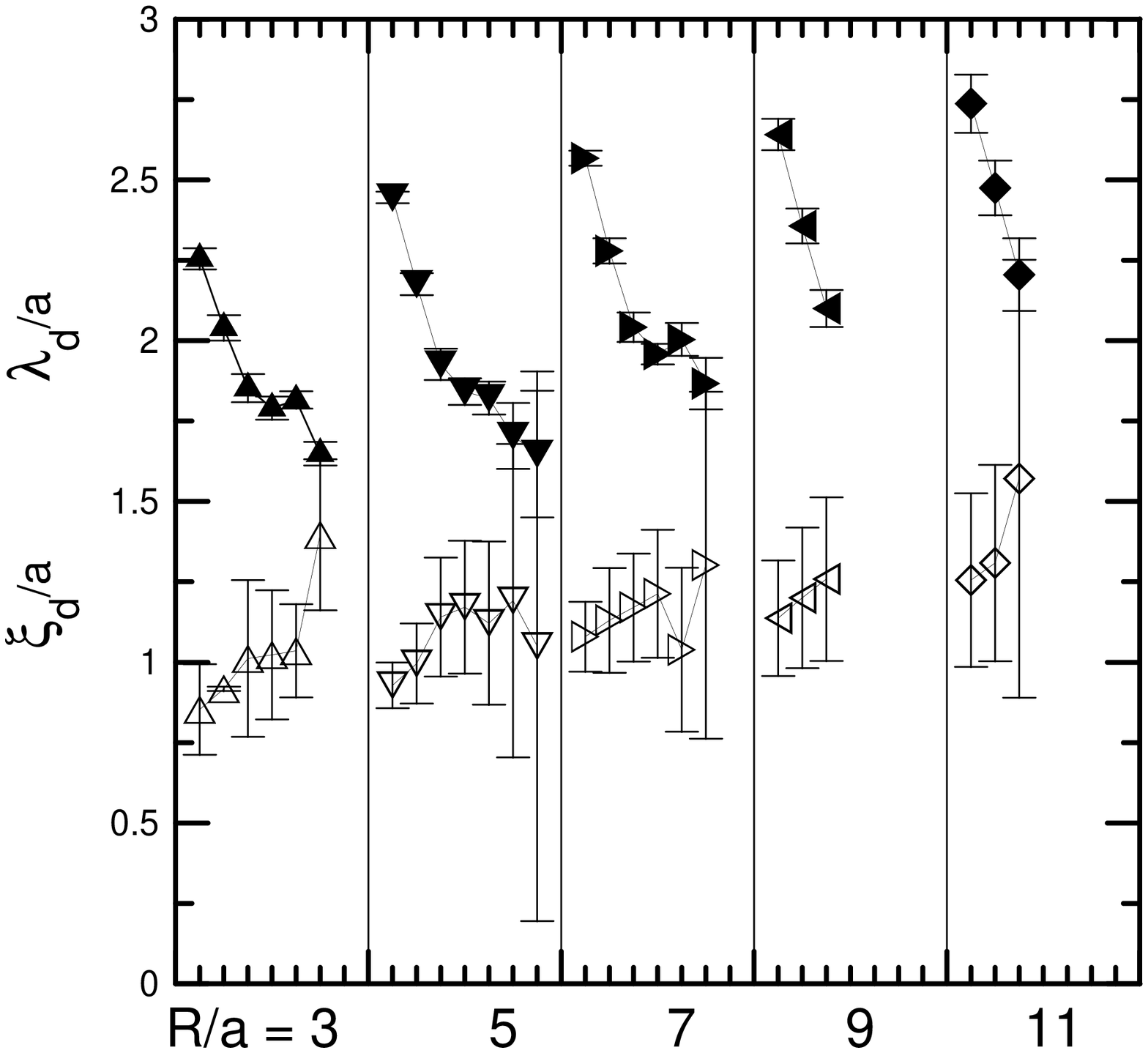}
\vskip -1.5in
\end{center}
\caption{Fitted parameters $\lambda_d/a$ and $\xi_d/a$ based on Eqs. (\ref{katrina}) and (\ref{rita}). 
 Within each group labeled by $R/a$, the $7$  ticks correspond to $T/a = 3,4, \ldots, 9$. }
\label{mars2}
\end{figure}
\begin{figure}[h]
\begin{center}
\vskip 1.0in
\includegraphics[trim=20mm 0 0 0,scale=0.4]{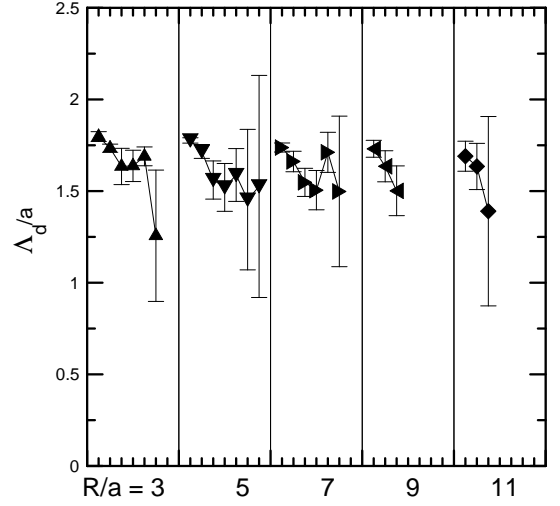}
\vskip -1.5in
\end{center}
\caption{Fitted parameters $\Lambda_d/a$ based on Eqs. (\ref{rita}). 
 Within each group labeled by $R/a$, the $7$  ticks correspond to $T/a = 3,4, \ldots, 9$. }
\label{mars3}
\end{figure}
\begin{figure}[h]
\begin{center}
\vskip 1.0in
\includegraphics[trim=20mm 0 0 0,scale=0.4]{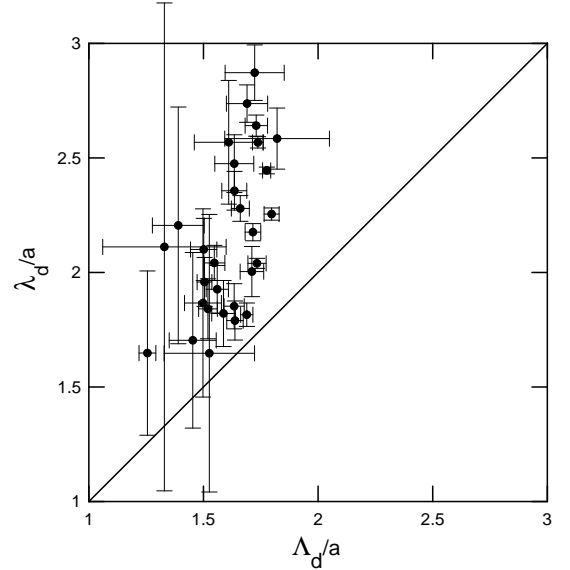}
\vskip -1.5in
\end{center}
\caption{Scatterplot of $\lambda_d/a$  vs. $\Lambda_d/a$. }
\label{mars4}
\end{figure}
Surveying the quality of the data for all $42$ data sets we arrived at the interval to identify the tails.  We chose 
the $10$ points: $r_i/a = 3.5, 4.0, \cdots, 8.0$ for $\chi^2$ fits.

For the flux, we search for the best values of $A$ and $\lambda_d$ 
\be
\int_{r' > r_i }  a^2 E_z   \;\; da'  &=& A \sqrt{r_i/a}e^{- r_i/\lambda_d}.
\label{katrina}
\ee
Figure \ref{mars1} shows the integrated  flux profile along with a $\chi^2$ fit.

In our analysis, we have used exponential tail forms appropriate for two dimensions.  The London relation in this case leads to exponential behavior 
$K_0(\zeta)$, see e.g. Ref.\cite{tinkham}, where $K_\nu$ is the modified Bessel function.  
We make use of the first term in the divergent asymptotic expansion
\ben
K_\nu(\zeta) &\approx& \sqrt{\frac{\pi}{2\zeta}} e^{-\zeta}\left( 1 + \frac{4 \nu^2-1}{8 \zeta}+ \cdots\right).
\een
The two dimensional choice requires roughly $r/a < R/2a$ which can not be achieved in the tail of the profile.  For example for $R/a=8 $ and a transverse distance $r/a=8$, the distance of the observation point to the sources is $8.9$ in lattice units, compared to a distance of $8.0$ from the vortex, which is clearly not a two dimensional problem.  Nevertheless it is reasonable at intermediate transverse distances and becomes correct as $R\rightarrow\infty$ and for large transverse distances.   And it allows a close comparison with Bali et al.\cite{bss}

Since we use expressions for the profile functions that have been integrated over the area, consider
\ben
\int_\eta^\infty \zeta^{-\nu} K_{\nu+1}(\zeta) d \zeta &=& \eta^{-\nu}K_\nu(\eta), \quad \lab{for}\; \nu = -1
\quad \lab{and} \\
K_{\nu-1}(\zeta) - K_{\nu+1}(\zeta) &=&\frac{2\nu}{z}K_\nu(\zeta), \quad \lab{for}\; \nu = 0.
\een
Hence 
\ben
\int_\eta^\infty  K_0(\zeta)\zeta d \zeta &=& \eta K_1(\eta)  \approx   \sqrt{\frac{\pi \eta}{2}}e^{-\eta}.
\een
As explained above the choice of integrating the flux reduces noise in the signal.

Next we do a $\chi^2$ fit of the fluxoid in a similar fashion
\be
\int_{r' > r_i } \left\{ a^2 E_z  + \frac{\Lambda^2_d}{a^2} \left(a^4 \lab{curl} \mb{J} ^{(m)}\right)_z  
\right\} \;\; da' =&& \nonumber \\ B \sqrt{r_i/a} e^{- r_i/\xi_d}.&&
\label{rita}
\ee
determining $\Lambda_d$, $B$, and $\xi_d$.   Figure \ref{mars1} also shows the sub-leading behavior after canceling the leading large $r$ behavior between $E_z$ and $(\lab{curl} \mb{J} ^{(m)})_z $, i.e. the fluxoid profile.  Also shown is 
the fit.

This demonstrates the heart of our method.  Through the fit we find a proportionality constant such that the London relation is satisfied asymptotically by the penetrating tails of the flux and the curl of magnetic current respectively.  Further the violation of the London relation is characterized by a second exponential at a shorter length scale.  Even without a fit, the data points alone support the cancellation  inherent in the construction.  For specificity, we chose this operational definition of $\xi_d$ with the understanding that it sets the scale but there can be a factor close to one relating it to the Ginzburg-Landau definition [49]. Further the borderline between type I and type II also has a factor close to one [51].  

{\em Using the parameters $\lambda_d$ and $\xi_d$ specifically defined by Eqs.(\ref{katrina}) and (\ref{rita}) then  the  condition Eq,(\ref{pogo}), $\kappa = \lambda_d/\xi_d > 1$ unequivocally signifies type II behavior.  Under that condition, there exists a region, possibly asymptotically, where the London relation is satified and that eliminates type I as a possibility.}

Out of the $42\times 2$ cases,  $\chi^2/\lab{d.f.}  < 1.0$ in the fits except for a few cases at small 
$R/a$ and small $T/a$.  The results of the  fits are shown in Figs. \ref{mars2}, \ref{mars3} and  \ref{mars4}.  

Note first in 
Figs. \ref{mars2} and \ref{mars3} that there are less than $42$ data points for each quantity.   This is because we set an inclusion criterion $\lambda_d \ge 1.1 \xi_d $.    The reason is that the method of isolating the two exponential forms breaks down as the parameters approach each other.  When they are close, large errors develop in trying to separate the two behaviors.  Hence the 
$10\%$ cut.  The absence of a data point indicates we tried and failed to produce a fluxoid which indicates that the parameters appear to be driven close to or into the type I regime.
Therefore every value of $R/a$ and $T/a$ with a data point indicates that we are able to separate the leading and subleading exponential forms which implies unambiguously type II behavior.

For $R/a = 5 \lab{ and } 7$  and for all values of $T/a$ except one, Fig. \ref{mars2} shows that we were able to construct  a fluxoid with the pattern shown in Fig. \ref{mars1}.  Whether one accepts our parameterization of the tails or not, this is strong model independent evidence that for these cases, the system behaves like type II.  On the other hand  for the values $R/a = 6$  and $T/a = 6$, fitting to the dual GLH model, Koma et al.\cite{kkis}, find type I behavior.   Because of differences in the number of smearing steps and the value of the smearing paramenter, we should not compare the two results for individual values of $T/a$.  However by calculating at many values of $T/a$  and bracketing their value of $R/a$ we are sidestepping that problem in a comparison.   

This conflict could probably be checked out without any fitting of profiles or any commitment to a particular method of interpreting the data.  Simply take the data points of the curl of the magnetic current profile and multiply by an arbitrary constant $\Lambda_d$ and scan its value looking for cancellation with the data points of the electric profile. A log plot will reveal if the search turns up a subleading 
exponential in the difference data points.  If so it can only mean type II, if not then it is borderline or type I.   

We have also repeated the whole analysis for a generic exponential fit to the tails in Eqs. (\ref{katrina},\ref{rita}) without regard to the power behavior multiplying the exponential, i.e. without the two-dimensional factor $\sqrt{r/a}$.  The pattern of  results for $\lambda_d$  and $\xi_d$ were quite similar to those in Fig. \ref{mars2} but the values of the two parameters  tended to be larger.   The present choice gave smaller errors and a more coherent 
scatterplot, Fig. \ref{mars4}.

\subsubsection{Comparison to Bali et al.\cite{bss}}

Our results allow a rather direct comparison with Bali et al.\cite{bss}.  It is interesting to compare their
methods and best estimate of $\lambda_d/a$ to ours. They chose $T/a = 6$, $R/a = 8$. Here are their steps in order:

\begin{tabbing}
(1) \= London relation in the tail of the flux, $\lambda_d/a = 1.82(7) $. \\ 

(2) \> Include physics of dual GLH  \\

 \>(a) but fitting only the flux,  $\lambda_d/a = 1.84(8) $.\\

 \>(b) Including the magnetic current, $\lambda_d/a = 1.99(5) $.\\

 \>(c) An alternative to (b), $\lambda_d/a = 1.62(2) $.\\

(3) \>  Best estimate: $\lambda_d/a =  1.84^{+20}_{-24}$.
\end{tabbing}
We regard step (1) as an essentially  correct definition of the penetration depth and with minor differences their methods agrees with ours\cite{footnote6}.  Our value e.g. $\lambda_d/a = 1.96(3)$ is for $T/a = 6$, $R/a = 7$ .  Koma et al.\cite{kkis} 
find $\lambda_d = 2.22(1)$ for $T/a = 6$, $R/a = 6$.

Step (2a) is a refinement acknowledging that there may be a non-zero coherence length detectable at intermediate distances in the tail.  

In the succeeding steps  (2b), (2c) and (3)  we differ since they rely on the dual GLH model. 
The fitted parameter  representing the penetration depth was found to take two different values statistically inconsistent with the first determination and each other when fitting two different ways.  A similar inconsistency would be expected based on the thesis of this paper if one tailored the fit to favor the $\Lambda_d$ definition of penetration depth vs. one that favors the $\lambda_d$ definition.  The quoted systematic error in the parameter was triple the original statistical error indicating to us trouble since the first approach was correct in our opinion. They attributed the systematic error to lattice artifacts coming from the fitting t definition.  In our opinion we regard this as evidence for problems with the dual GLH model interpretation of the data. 

We may or may not differ on type I vs. type II. Bracketing their value $T/a = 6$, $R/a = 8$ we find type II 
for the same time extent and $R/a=7$, and our method breaks down for our next value $R/a=9$.  They find type I.   Given this finding, ironically it negates their starting point, step (1), in using $K_0(r/\lambda_d)$ since that form is not applicable in the case of type I.  (It is not clear to us if the succeeding steps depend on step (1).)

\subsubsection{The effect of Gribov copies}

There is more we can say.  Bali et al.\cite{bss} used the methods of  Bornyakov et al.\cite{bbms} to find the best maximum of the gauge fixing algorithm.  This allows the use of successive estimates to look for gauge copy dependencies in results.  On the other hand we did a simple overrelaxation gauge fixing algorithm.    
The question arises whether this difference of our methods with Bali et al. (step(1)) could lead to differing results in fitting parameters. It turns out that the results are insensitive to the maximum selection criteria of Ref.\cite{bbms}.  To see this consider the upper left hand graph in Fig. 4 of Koma et al.\cite{kkis}.  We see that the flux profile is remarkably insensitive to Gribov copies. There is a very small Gribov copy dependence in the core of the vortex and it appears to entirely go away in the tail in the range where we and Bali et al. determined the tail, $r/a \ge 3.5$. If we were to take the data from Ref. \cite{kkis} and use our model independent methods we would obviously have to  find our $\lambda_d$ parameter insensitive to copies as would Bali et al.\cite{bss} (in their steps (1) and (2a), using the designations in the previous subsection).

There is another parameter we call $\Lambda_d$, the square of which is the proportionality factor between the tails of the flux and the curl of the magnetic current.  The upper right hand graph in the same figure, Fig.4, Ref.\cite{kkis},  shows there is detectable but small Gribov copy dependence for the magnetic current in the tail, for $r/a \ge 3.5$ where we work.  Using their data and our analysis, that implies $\Lambda_d$ could have a small Gribov copy dependence.  But if their analysis gives a penetration depth with measurable copy dependence  then it must come from an artifact of the model since two independent parameters,  $\Lambda_d$ and $\lambda_d$ are forced to be equal by the model.

If the dual GLH model were the correct model of the data then the fitted parameters would have to give a faithful representation of the physical observables apparent in the tails.  And this would be a very strong test of the model if the parameters truly represented the tails.  
Unfortunately there are no log graphs of profiles in the literature where one could independently verify the fits by eyeball.

Next, consider the effect of copies on the coherence length.  Indeed our $\xi_d$ may very well have Gribov copy dependence, though small, for the same reason that $\Lambda_d$ might show this dependence.

That is not to say the the refinement of using the Ref.\cite{bbms} is not necessary.  For higher values of $\beta$, Koma et al.\cite{kkis} have shown that the Gribov copy dependence is more pronounced.

\subsubsection{Other features}

Some general features of Fig. \ref{mars2} are worth noting.   Increasing $T/a$ drives the $\lambda_d$ 
and $\xi_d$ together. With limited statistics we can not take $R/a$ large enough to make a definitive statement as to whether type II behavior persists.

The results for $\Lambda_d$  are shown in Fig. \ref{mars3}.   We also give a scatter plot of $\Lambda_d$ vs. $\lambda_d$, Fig. \ref{mars4}.  There is no clear tendency for them to be equal as they are in the dual GLH model.

\begin{table}[h]
\begin{center}
\begin{tabular}{c|ccc} \hline \hline 
 fitting range, $r/a$ & $[3.5,8]$ & $[2.5,8]$ & $[3.5,10]$ \\
 no. of pts.    & $10$   & $12$ & $14$ \\ \hline \hline
 $\lambda_d/a$ & $1.96(3)$  & $1.92(2)$ & $2.00(3)$ \\
 $\chi^2/\lab{d.f.} $  & $0.40$ & $0.56$& $0.63$ \\ \hline \hline
 $\xi_d/a$   & $1.21(20)$  & $1.28(11)$ & $1.24(16)$ \\
 $\Lambda_d/a$ & $1.50(11)$  & $1.47(9)$ & $1.49(8)$ \\
 $\chi^2/\lab{d.f.} $  & $0.03$ & $0.06$ & $0.14$ \\ \hline \hline
 $\Phi_{SU(2)}$  && 1.72(3)& \\
 $\chi^2/\lab{d.f.} $ && 1.0 & \\ \hline \hline
\end{tabular}
\end{center}
\caption{First data column: parameters determined from  $T/a=6$, $R/a=7$ using standard interval. Second data column: sensitivity to lower cut. Third data column: sensitivity to upper cut.  The value of $\Phi_{SU(2)}$ does not involve cuts.}
\end{table}

In Table II we show the sensitivity of the parameters  for a particular Wilson loop size to varying the fitting interval.
Note that the second row of values of $\chi^2/df$ for the fluxoid fit are exceptionally small.  From Fig. \ref{mars1} we see that the fit to the points of the fluxoid is excellent but the error bars appear to be too large given the smooth behavior of the data points.  We attribute this to error propagation in forming the fluxoid where there might be  correlations in the two canceling contributions.

\subsection{Electric currents}

\begin{figure}
\begin{center}
\vskip 1.0in
\includegraphics[trim=30mm 0 0 0 ,scale=0.4]{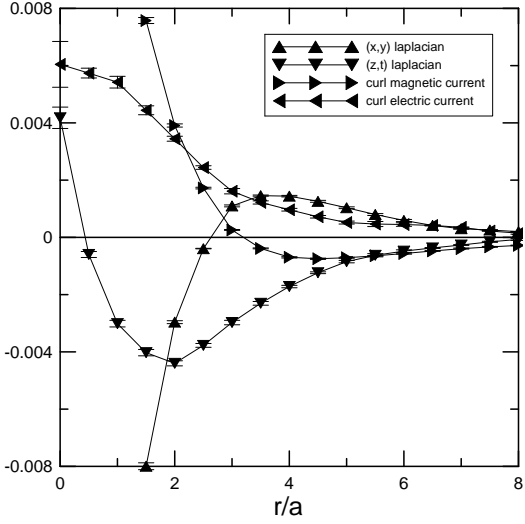}
\vskip -1.5in
\end{center}
\caption{The four contributions to the identity, Eq.(\ref{stove}). The sum of the four graphs vanishes identically.
 $R/a = 7$, $T/a = 3$.
 }
\label{mercury1}
\end{figure}
\begin{figure}
\begin{center}
\vskip 1.0in
\includegraphics[trim=30mm 0 0 0,scale=0.4]{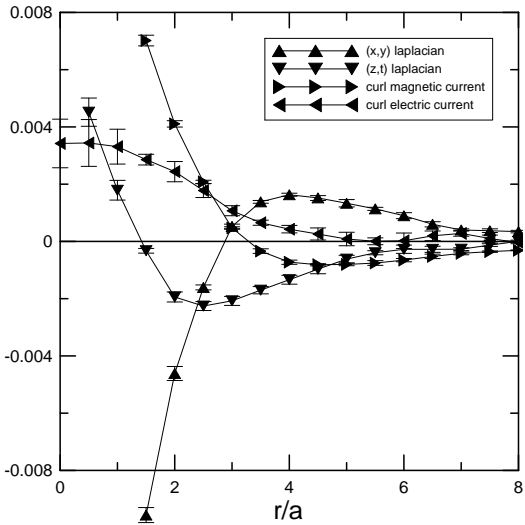}
\vskip -1.5in
\end{center}
\caption{Same as Fig. \ref{mercury1} except $T/a = 6$}
\label{mercury2}
\end{figure}
\begin{figure}
\begin{center}
\vskip 1.0in
\includegraphics[trim=30mm 0 0 0,scale=0.4]{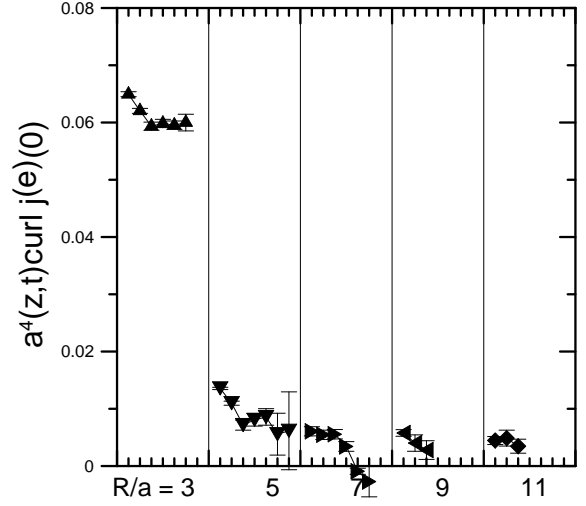}
\vskip -1.5in
\end{center}
\caption{The $(z,t)$ curl of the electric current evaluated on the vortex axis. 
 Within each group labeled by $R/a$, the $7$  ticks correspond to $T/a = 3,4, \ldots, 9$.}
\label{mercury3}
\end{figure}
Figure \ref{mercury1} shows an example of the effect of electric currents in this problem.   The four graphs correspond
to the four terms in the identity (a rewriting of Eq.(\ref{steve}) in dimensionless form and supressing the
expectation value bracket):
\be
0 = a^3
\left(
\Delta^-_1   J^{(m)}_2 -  \Delta^-_2   J^{(m)}_1 \right) - 
a^2\left(\Delta^-_1\Delta^+_1 + \Delta^-_2\Delta^+_2 \right)  F^{(3)}_{43} 
 &&  \nonumber \\
+a^3
\left( \Delta^+_3 J^{(e)}_4
  -  \Delta^+_4  J^{(e)}_3 \right) 
  - a^2\left(\Delta^-_3\Delta^+_3 + \Delta^-_4\Delta^+_4    \right) F^{(3)}_{43}. \quad&&
\label{stove}
\ee
The curl of the magnetic current is in the $(x,y)$ plane and the curl of the electric current is in the $(z,t)$ plane. All four terms live on the same plaquette, which is the same as the $E_z$ plaquette.

In a continuum dual GLH model, the third term, the electric current term, vanishes. Further if we consider an infinitely long static vortex in the dual GLH model, then the fourth term also vanishes.  

Both the third and fourth terms are significant and remain so over a wide range of $R$'s and $T$'s presented here.  The discrepancy between $\Lambda_d$ and $\lambda_d$ further support the presence of the third and fourth terms over this range.  

As we go to larger values of $T/a$ the behavior of the electric currents is  more complicated  as shown in  Fig. \ref{mercury2}.  In particular there are wiggles in the electric current term in the tail.  There are examples of sign changes for other values of $R/a$ and $T/a$.  It is beyond the scope of this paper to delve into such features since we do not have a well motivated model of these currents.  Rather we wish to point out here the evidence that the electric current persists for the range of $R/a$'s and $T/a$'s studied here.   

Fig. \ref{mercury3} shows the value of this term on the axis.  For $R/a = 3$ the curl term on axis picks up the large antiscreening contribution adjacent to the source which is discussed elsewhere\cite{bdh,dhh,bss}.  At $R/a = 5$ and larger, this value drops.  But it persists and can change sign.  

The $(z,t)$ laplacian term also persists.  It measures how well we are in the regime of large $R/a$ and $T/a$. The non-vanishing of this term contributes to the non-equality of  $\lambda_d$ and $\Lambda_d$.   Such a term will arise in the three dimensional dual GLH model used by Koma et al.\cite{kkis}.  It would be interesting to see how well it fits this term.  From our point of view we need a three dimensional model that does not commit one to the dual GLH model if we are to resolve the issues raised here.

If one were to drop all expectations about dual superconductivity and simply ask in an unbiased way how well the dual GLH model fits the simulation data, then one ought to do a $\chi^2$  fit that includes {\em all four} of these terms.  They all live on the same plaquette.  Then one would be truly testing the validity of the dual GLH model.   Of course the result would be that the contribution  to $\chi^2$ involving the electric current would diverge as statistics improved.
And the contribution involving the $(z,t)$ Laplacian would diverge if the source in the model  did not match the source in the simulation.   Measuring the latter term offers a good test of how well one matches the idealization of an infinitely long static string.

\subsection{Truncated monopole loops}
\begin{figure}[h]
\begin{center}
\vskip 0.7in
\includegraphics[trim=20mm 0 0 0,scale=0.4]{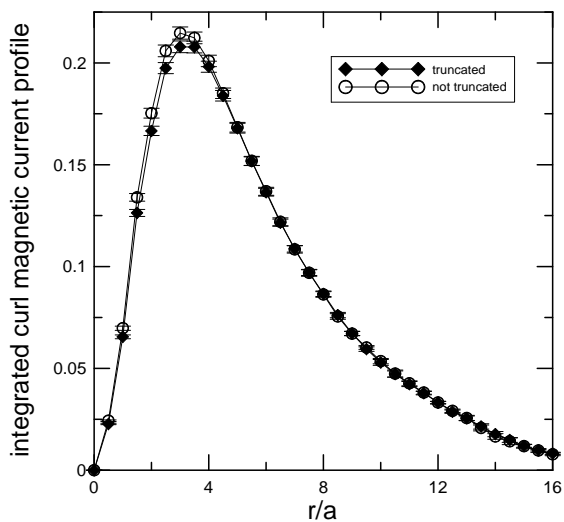}
\vskip -1.1in
\end{center}
\caption{The effect of truncation of monopole loops, keeping only the percolating cluster.}
\label{saturn}
\end{figure}
Finally we give a result that is disjoint from the body of this paper.   We looked at monopole loops using the standard DT definition, i.e. constructed using $F^{(1)}_{\mu \nu}$.  In this case there is a distinction between connected and disconnected loops.   Earlier studies have shown that there is a single large connected percolating cluster of monopole
currents that dominates confinement physics\cite{ht,bcgp,cz,ciks}.  There is a sharp distinction between this percolating cluster and the large number of very small loops.   

We confirm this result in our configurations and find that the big cluster
contains  $ \sim 40\%$ of the current.  We compared our calculations of the current with the truncated version, dropping 
$ \sim 60\%$ of the contribution to the current coming from the small loops.  Our results are shown in Fig. \ref{saturn}. This gives the integrated curl $ \int_{r' < r } (\mb{\lab{curl} \mb{J} }  )\cdot d \mb{a'} $ as a function of $r$.  The truncation has  a very minor effect in the core and essentially no effect in the tail which means that it has a minor effect in determining the parameters of the vortex.

\section{Summary and Conclusions}

We regard the most interesting open question in studies of the dual Abrikosov vortex is whether the system 
behaves as type I, type II or on the border line.   We have cast doubt on the use of the dual GLH model as a reliable way to analyze simulation data. There are no dynamical electric currents in the dual GLH model for the same reason there are no dynamical magnetic currents in the GLH model.  Hence it is wrong, but how wrong?  We argue that these currents are important enough to cast doubt on the prior efforts to establish the type of superconductivity. 

How reliable is our model independent approach in trying to answer the question?  The definition of the GL coherence length $\xi$ and the GL borderline parameter $\kappa \approx \lambda/\xi$ are not easily divorced from the GL theory. 
And we see in real world  superconductivity there are various ways to set the coherence length scale, but they are related factors close to 1, see Tinkham\cite{tinkham} and [49,51].  We have sidestepped these issues that involve the dual GLH model and instead defined the parameters based directly on the behavior of the flux and fluxoid, See Sec. III C .

Lets put off that question and try to answer a simplier question.  What is the apparant type I/II behvior when the system is probed by a specfic Wilson loop?  	We have proposed a way to answer it. 

Consider Fig. \ref{mars1}.  We arrived at the fluxoid profile by finding a value of $\Lambda_d$ such that $\Lambda^2_d \times$ curl of the magnetic current cancels against the electric flux in the sense that the difference has an exponential behavior with a shorter characteristic length.  Ignore the fits in the graph and the details of the functional forms.  One can arrive at this just by stepping the value of $\Lambda_d$ until this value is found.  It doesn't even require determining $\lambda_d$ or $\xi_d$. 

The data points alone in Fig. \ref{mars1},  makes it very clear that 
the London equation is satisfied asymptotically for large $r/a$.  This is unequivocal type II behavior for this 
$R/a=7, T/a=6$ Wilson loop source.   The London equation is irrelevant to type I behavior because the ramp-up of the order parameter occurs over a larger distance than the penetration of the flux.

Fig. \ref{mars2} shows $25$ Wilson loop sizes for which that construction was successful.  There may be more but we made a conservative cut in the data as described in Sec. III C.  There is clearly room for improvement in our procedure to identify the fluxoid.  As we go to larger loop sizes and better suppression of excited states, the system seems to be moving toward  type I behavior.

Koma et al.\cite{kkis}  reported  type I behavior for  $R/a=6, T/a=6$.  We report $13$ loop sizes that bracket their loop size for which we are successful in the fluxoid construction. So we would surmise that had we calculated 
$R/a=6$ we would most likely report type II for all our values of $T/a$ with the largest value possibly in question.
Bali et al.\cite{bss} reported type I or  $R/a=8, T/a=6$. We did not calculate this case or bracket it well enough to speculate what we would find.  

Gubarev et al.\cite{gips} reported that the system may lie on the boundary.  This would clearly be the most satisfying  outcome theoretically since it implies there may be something universal happening. 

Given that the total flux in the vortex and the shape of the vortex are intimately connected to the electric and magnetic Maxwell equations, we considered it a high priority to use a definition of flux that guarantees the equations are satisfied for lattice averages.  A bonus was an identity relating the magnetic and electric currents which was exact only for our choice of definitions of flux. This identity, Eq.(\ref{stove}) was central in elucidating the problems associated with the dual GLH model.

\begin{acknowledgments}
Work supported in part by the U.S. Department of Energy under grant no. DE-FG05-01 ER 40617.
Our numerical computations were performed in part on the Louisiana State University Intel 
Xeon cluster, SuperMike, that is operated by LSU's Center for Computation and Technology (CCT).
\end{acknowledgments}

\appendix


\begin{thebibliography}{0}

\bibitem{suzuki} T. Suzuki, Nucl. Phys. B (Proc. Suppl.) {\bf 30}, 176 (1993).
\bibitem{cp} M. N. Chernodub and M. I. Polikarpov, ``Confinement, duality and nonperturbative aspects of QCD", Ed by P. van Baal, Plenum Press, p. 387, hep-th/9710205.
\bibitem{haymaker1} R. W. Haymaker
 Lectures, Int. Sch of Physics, "Enrico Fermi", Varenna, Italy, 1995,
"Varenna 1995, Selected topics in nonperturbative QCD" 175-201.  hep-lat/9510035.
\bibitem{polikarpov} M.I. Polikarpov, Nucl. Phys. (Proc. Suppl.) {\bf 53}, 134   (1997)
\bibitem{digiacomo} A. Di Giacomo, Prog. Theor. Phys. Suppl. {\bf 131}, 161  (1998), hep-lat/9802008. 
\bibitem{cgpv} M.N.Chernodub, F.V.Gubarev, M.I.Polikarpov, and A.I.Veselov,
1997 Yukawa Int. Seminar on "Non-perturbative QCD - Structure of QCD Vacuum -" (YKIS'97), Kyoto, 
Prog. Theor. Phys. Suppl. {\bf 131}, 309 (1998), hep-lat/9802036.
\bibitem{haymaker2} R. W. Haymaker, Phys. Rep. {\bf 315}, 153 (1999).
\bibitem{cgpz} M.N.Chernodub, F.V.Gubarev, M.I.Polikarpov, and V.I.Zakharov, hep-lat/0103033.
\bibitem{ripka} G. Ripka, Trento Lectures, 2003, Springer, Berlin (2004), 639, hep-ph/0310102.
%
%
\bibitem{klsw} A. S. Kronfeld, M. Laursen, G. Schierholz, and U.-J. Wiese, Phys. Lett. B {\bf 198}, 516 (1987).
\bibitem{sy} T. Suzuki and I. Yotsuyanagi, Phys. Rev. D {\bf 42}, 4257 (1990).
%
%
\bibitem{snw} J. D. Stack, S. D. Neiman, and R. J. Wensley, Phys. Rev. D {\bf 50}, 3399 (1994).
\bibitem{ss} H. Shiba and T. Suzuki, Phys. Lett.  B {\bf 333}, 461 (1994).
%
%
\bibitem{ht} A. Hart and M. Teper, Phys. Rev. D {\bf 58},  014504 (1998).
\bibitem{bcgp} V.G. Bornyakov, M.N. Chernodub, F.V. Gubarev, M. I. Polikarpov, 
T. Suzuki, a.I. Veselov, and V.I. Zakharov, Phys. Lett. B {\bf 537},  291 (2002), hep-lat/0103032.
\bibitem{cz} M.N. Chernodub and V.I. Zakharov, Nucl.Phys. B {\bf 669}, 233 (2003).
\bibitem{ciks} M.N.Chernodub, Katsuya Ishiguro, Katsuya Kobayashi, and Tsuneo Suzuki, 
Phys. Rev. D {\bf 69}, 014509 (2004), hep-lat/0306001.
%
%
\bibitem{pisa3} L. Del Debbio, Di Giacomo, and G. Paffuti, Phys. Lett. B, {\bf 349}, 513 (1995).
\bibitem{pisa2} A. Di Giacomo, B. Lucini, L. Montesi, and G. Paffuti,
Phys. Rev. D {\bf 61}, 034503 (2000); 034504 (2000).
\bibitem{pisa1} J.M. Carmona, M. D'Elia, A. Di Giacomo, B. Lucini, and G. Paffuti,
Phys. Rev. D {\bf 64}, 114507 (2001).
%
\bibitem{suganuma1} H. Suganuma, K. Amemiya, H. Ichie, H. Matsufuru, Y. Nemoto, and T.T. Takahashi, 
(Confinement 2000), Osaka Japan, hep-lat/0407020.
\bibitem{suganuma4} H. Suganuma, K. Amemiya, and H. Ichie,  Nucl. Phys. 
(Proc. Suppl) {\bf 83}, 547 (2000), hep-lat/0407015.
%
\bibitem{sbh} V. Singh, D. A. Browne, and R. W. Haymaker, Phys. Lett. B {\bf 306}, 115 (1993).
\bibitem{mes} Y. Matsubara, S. Ejiri, and T. Suzuki, Nucl. Phys. B (Proc. Suppl.), {\bf 34}, 176 (1994).
\bibitem{ph} Y. Peng  and R. W. Haymaker, Phys. Rev. D {\bf 52}, 3030 (1995).
\bibitem{cc} P. Cea and L. Cosmai, Phys. Rev. D {\bf 52}, 5152 (1995).
\bibitem{bss} G. S. Bali, C. Schlichter, and K. Schilling, 
Prog. Theor. Phys. Suppl. {\bf 131}, 645 (1998).
\bibitem{bali} G. S. Bali, Talk 3rd International Conference on Quark Confinement and the Hadron Spectrum (Confinement III), Newport News, VA, 7-12 Jun 1998, hep-ph/9809351.
\bibitem{gips} F. V. Gubarev, E.-M Ilgenfritz, M. I. Polikarpov, and T. Suzuki,  
Phys. Lett. B {\bf 468}, 134 (1999).
\bibitem{kkisp} Y. Koma, M. Koma, E.-M. Ilgenfritz, T. Suzuki, and M.I. Polikarpov, Phys. Rev. D {\bf 68},
094018 (2003), hep-lat/0302006.
\bibitem{kkis} Y. Koma, M. Koma, E.-M. Ilgenfritz, and T. Suzuki, Phys. Rev. D {\bf 68},
114504 (2003), hep-lat/0308008.
\bibitem{bcp} V. A. Belavin, M. N. Chernodub, and M. I. Polikarpov, hep-lat/0403013.
\bibitem{mioys} Y. Matsubara, S. Ilyar, T. Okude, K. Yotsuji, and T. Suzuki, Nucl. Phys. (Proc. Suppl.), {\bf 42}, 
529 (1995).
\bibitem{ekmosy} S. Ejiri, S. I. Kitahara, Y. Matsubara, T. Okude, T Suzuki, and K. Yasuta, 
Nucl. Phys.  (Proc. Suppl.) {\bf 47}, 322 (1996).
\bibitem{sijs} A. J. van der Sijs, Nucl. Phys. (Proc. Suppl.) {\bf 73}, 548 (1999).
\bibitem{fkst} S. Fujimoto, S. Kato, T. Suzuki, and T. Tsunemi, Prog. Theor. Phys. Suppl. {\bf 138}, 36 (2000).
\bibitem{bbms}B. S. Bali, V. Bornyakov, M. M{\" u}ller-Preussker, and K. Schilling, Phys. Rev. D {\bf 54},
2863 (1996).
%
\bibitem{tinkham} M. Tinkham, \underline{Introduction to Superconductivity} McGraw Hill, New York, (1975).
%
\bibitem{footnote1} See Sec. 4-3, Ref.\cite{tinkham}.
%
%
\bibitem{huebener}  R. P. Huebener, \underline{Magnetic} \underline{flux} \underline{structures} \underline{in} \underline{supercon-} \underline{ductors}, Springer, New York 1979.
%
%
\bibitem{callaway}  D. A. Callaway, Nucl. Phys. {\bf B 344}, 627 (1990).
%
%
\bibitem{zfks} M. Zach, M. Faber, W. Kainz, and P. Skala  Phys. Lett.  B {\bf 358},  325 (1995).
\bibitem{dhh} G. DiCecio, A. Hart, and R. Haymaker,  Phys. Lett. B  {\bf 441}, 319 (1998).
\bibitem{dt} T. A. DeGrand and D. Toussaint, Phys. Rev. D {\bf 22},  2478 (1980).
%
%
\bibitem{mh}   R.W. Haymaker and T. Matsuki, Ed. by Suganuma et al, Wako 2003, Color Confinement and Hadrons in
Quantum Chromodynamics, 21-24 July, 2003 World Scientific, Singapore, 2004, pp.60-71,
Lattice 2003, Tsukuba,  Nucl. Phys. Proc. Suppl. {\bf 129}, 641 (2004), hep-lat/0310025.
%


\bibitem{footnote2}In the superconducting literature, fluxoid refers to the integral of the fluxoid density over the profile. See Ref.\cite{tinkham}, Eq.(4.42). To be more correct we should use the term fluxoid density profile but then to be consistent we would also need to use the term flux density profile.



\bibitem{footnote3}In Ginzburg Landau theory, the definition of $\xi$ can be taken from Eq. (4.18),  where the order parameter $f = \lab{tanh} \frac{\nu r}{\xi}$ with $\nu \approx 1$ Eq. (5.8) \cite{tinkham}.  Therefore the scale of the ramp-up of the order parameter is goverened by $\xi$ but there may be a correction factor. Our specific definition of $\xi_d$ is given by Eq.(\ref{rita}) and described in the following paragraph.  We discuss this further in the summary.

\bibitem{footnote4}An alternative normalization is found in the superconducting  literature\cite{tinkham}  and in some cases the dual superconductivity literature including 
Bali et al.\cite{bss} has the coherence length equal to $\xi/\sqrt{2}$. In this case $\kappa < 1/\sqrt{2}$ for type I and $\kappa > 1/\sqrt{2}$ for type II.  

\bibitem{footnote4a}There can be a `fudge factor' in the definition of 
$\kappa$, Eq.(4.27)\cite{tinkham}.


\bibitem{footnote5}Quantities
with a `hat'  mean those which appear in the lattice simulation
without appending factors of the gauge coupling constant $e$ or $g$ or the lattice spacing $a$.

%
\bibitem{bdh}      K. Bernstein, G. DiCecio and R. W. Haymaker, Phys. Rev. {\bf D55}, 6730 (1997).



\bibitem{poulis} G. Poulis,  Phys. Rev. D {\bf 54},  6974 (1996).


\bibitem{footnote6}They use the Bessel function $K_0(\zeta)$ which is appropriate for a  two dimensional geometry of an infinitely long vortex. We use the asymptotic expansion of the $K_\nu(\zeta)$. They use it out to $r/a = 7.07$. the static sources are at a lattice distance of 8.12 so it is not two dimensional. Everyone faces this problem one way or another. It may be less of a problem for  Koma et al.\cite{kkis} since they solve a 3-d lattice model but thereby 
commit to the dual GLH model.


\bibitem{shb} V. Singh, R. W. Haymaker and D. Browne  Phys. Rev. D {\bf 47},  1715 (1993).


\end{thebibliography}
\end{document}